\documentclass{article}
\usepackage{fullpage}

\usepackage[ruled,boxed,lined,linesnumbered]{algorithm2e}
\usepackage{graphicx}
\usepackage{amsmath,amssymb,amsfonts,amsthm}
\usepackage{enumerate}
\usepackage{color,multirow}
\usepackage{caption}
\usepackage{subcaption}
\usepackage{moreverb}
\usepackage{todonotes}
\usepackage{hhline}
\usepackage[bookmarks=false]{hyperref}
\usepackage{multirow}
\usepackage{listings}
\usepackage{color}
\newenvironment{remark}[1][Remark]{\begin{trivlist}
\item[\hskip \labelsep {\bfseries #1}]}{\end{trivlist}}
\newcommand{\ignore}[1]{}
\usepackage{booktabs}
\usepackage{tikz}
\newtheorem{mydef}{Definition}

\tikzstyle{cnode} = [draw, circle,scale=0.7]
\tikzstyle{level 1} = [level distance=.19\textwidth, sibling distance=.4\textwidth]
\tikzstyle{level 2} = [sibling distance=.2\textwidth]

\newcommand{\tm}{\textsuperscript{\textregistered}}

\newcommand{\extadd}{\mathbin{\hspace{0.65em}\text{\makebox[0pt]{\resizebox{0.3em}{0.85em}{\(\updownarrow\)}}\raisebox{0.15em}{\makebox[0pt]{\resizebox{1.3em}{0.2em}{\(\leftrightarrow\)}}}}\hspace{0.65em}}}
\newcommand{\skinnyextadd}{\mathbin{\hspace{0.65em}\text{\makebox[0pt]{\resizebox{0.3em}{0.85em}{\(\updownarrow\)}}\raisebox{0.15em}{\makebox[0pt]{\resizebox{1.3em}{0.2em}{$-$}}}}\hspace{0.65em}}}

\newcommand{\mytitle}{An efficient multi-core implementation of a
  novel HSS-structured multifrontal solver using randomized sampling}

\newcommand{\myshorttitle}{Randomized sampling HSS-structured sparse solver}
\title{\mytitle{}}

\author{Pieter Ghysels\footnotemark[1] \and
  Xiaoye S. Li\footnotemark[1] \and
  Fran\c{c}ois-Henry Rouet\footnotemark[1] \and\newline
  Samuel Williams\footnotemark[1] \and
  Artem Napov\footnotemark[2]
}

\begin{document}
\maketitle
%\slugger{sisc}{xxxx}{xx}{x}{x--x}%slugger should be set to mms, siap, sicomp, sicon, sidma, sima, simax, sinum, siopt, sisc, or sirev

\renewcommand{\thefootnote}{\fnsymbol{footnote}}
\footnotetext[1]{Lawrence Berkeley National Laboratory, Berkeley, CA
  94720, USA. ({\tt \{pghysels,xsli,fhrouet,swwilliams\}@lbl.gov})}
\footnotetext[2]{Universit\'e Libre de Bruxelles, B-1050 Brussels, Belgium.}
\renewcommand{\thefootnote}{\arabic{footnote}}

\begin{abstract}
  % We present an efficient implementation of a sparse direct solver,
  % based on the multifrontal algorithm, that exploits low-rank
  % compression of the dense frontal matrices that arise in the
  % factorization phase.
  We present a sparse linear system solver that is based on a
  multifrontal variant of Gaussian elimination, and exploits low-rank
  approximation of the resulting dense frontal matrices.
  % The observation that for many engineering applications the
  % frontal matrices have low-rank off-diagonal blocks motivates the
  % choice of the hierarchically semiseparable (HSS) matrix format for
  % the frontal matrices.
  We use hierarchically semiseparable (HSS) matrices, which have
  low-rank off-diagonal blocks, to approximate the frontal
  matrices. For HSS matrix construction, a randomized sampling
  algorithm is used together with interpolative decompositions.
 % based on rank-revealing QR factorization.
  The combination of the
  randomized compression with a fast ULV HSS factorization leads to a
  solver with lower computational complexity than the standard
  multifrontal method for many applications, resulting in speedups up
  to $7$ fold for problems in our test suite.
 % The method can be used as a direct solver or as a
 % preconditioner, depending on the compression tolerance.
  The implementation targets many-core systems by using task parallelism
  with dynamic runtime scheduling. Numerical experiments show
  performance improvements over state-of-the-art sparse direct
  solvers. The implementation achieves high performance and good
  scalability on a range of modern shared memory parallel systems,
  including the Intel\tm{} Xeon Phi (MIC). The code is part of a
  software package called STRUMPACK -- STRUctured Matrices PACKage,
  which also has a distributed memory component for dense
  rank-structured matrices.
\end{abstract}

%\begin{keywords}sparse Gaussian elimination, multifrontal method, HSS matrices,
%  parallel algorithm\end{keywords}

%\begin{AMS}\end{AMS}

%\pagestyle{myheadings}
%\thispagestyle{plain}
\markboth{P. Ghysels et al.}{\myshorttitle{}}

\section{Introduction}
Solving large linear systems efficiently on modern hardware is an
important requirement for many engineering high performance computing
codes.  For a wide range of applications, like those using finite
element, finite difference or finite volume discretizations of partial
differential equations (PDEs), the resulting linear system is
extremely sparse. Fast solution methods exploit this sparsity, but
also arrange the computations in such a way that most of the
computational work is done on smaller dense submatrices. The reason
for this is that operations on dense matrices can be implemented very
efficiently on modern hardware. The multifrontal
method~\cite{duff1983multifrontal,liu1992multifrontal} is an example
of a sparse direct solver where most of the work is done on dense,
so-called frontal matrices. Unfortunately, dense linear algebra
operations, for instance LU decomposition, require $\mathcal{O}(N^3)$
operations, where $N$ is the matrix dimension. In a multifrontal
solver these dense operations end up being the bottleneck. However, it
has been observed that for many applications the dense frontal
matrices have some kind of low-rank
structure~\cite{chandrasekaran2010numerical}.

In~\cite{xia2013randomized}, a rank-structured multifrontal method is
presented in which the larger frontal matrices are approximated by
hierarchically semiseparable (HSS)~\cite{vandebril2005bibliography}
matrices. For certain model problems, this leads to a solver, or
preconditioner, with linear or almost linear complexity in the total
number of degrees of freedom in the sparse linear system. Here, we
present an efficient implementation of a slightly modified version of
the algorithm presented in~\cite{xia2013randomized}. The algorithm
in~\cite{xia2013randomized} handles only symmetric positive definite
systems, while the code presented here targets general non-symmetric
non-singular matrices. For HSS compression, a randomized sampling
algorithm from~\cite{martinsson2011fast} is used. Earlier HSS
construction methods, see~\cite{xia2010fast}, cost at least
$\mathcal{O}(N^2)$, whereas the randomized method in combination with
a fast matrix-vector product has a linear or almost linear complexity,
depending on the rank-structure of the frontal matrix.
\ignore{
\todo[inline]{[FH] The sentence above is true only if you have a fast
matvec or in our special sparse context (the sample is assembled
from HSS matvecs), but not in general.
I feel like the intro has a couple sentences about complexity that
are confusing because they mix sparse and dense and don't have enough
context. I would leave only the following: MF=\(N^2\), MF-HSS+RS=\(N\)-ish
\textbf{for model problems}.}
}

An important concept used in the randomized compression algorithm is
the interpolative or skeleton
decomposition~\cite{cheng2005compression}. Use of this decomposition
leads to a special structure of the HSS generator matrices (see
Eq.~\eqref{eq:UVstructure}). The HSS factorization used
in~\cite{xia2013randomized} for symmetric matrices, and
in~\cite{xia2012superfast} for non-symmetric matrices, exploits this
special structure in a so-called ULV-like decomposition. In this
paper, the ULV decomposition from~\cite{xia2012superfast} is used.

The HSS format is a subclass of a more general type of hierarchical
rank-structured matrices called $\mathcal{H}$-matrices
\cite{borm2003introduction}. HSS matrices are similar to
$\mathcal{H}^2$-matrices, another subclass of $\mathcal{H}$-matrices,
in the sense that both formats have the special property that the
generators are hierarchically nested (see Eq.~\eqref{eq:nestedUV} for
what this means for HSS). This is typically not the case in the more
general $\mathcal{H}$, the block low-rank
(BLR)~\cite{amestoy2014improving}, the sequentially semi-separable
(SSS)~\cite{vandebril2005bibliography} or the hierarchically
off-diagonal low-rank (HODLR)~\cite{SivaramPHD} formats (all of which
are $\mathcal{H}$-matrices). In HSS and HODLR only off-diagonal blocks
are approximated as low-rank whereas $\mathcal{H}$, $\mathcal{H}^2$
and BLR allow more freedom in the
partitioning. In~\cite{amestoy2014improving}, BLR is used to
approximate dense frontal matrices in the multifrontal solver
MUMPS~\cite{amestoy2001fully} while in other recent
work~\cite{AminfarAD14} HODLR has also been proposed to accelerate a
multifrontal solver.  Both HSS and HODLR use similar hierarchical
off-diagonal partitioning, but HSS further exploits the hierarchically
nested bases structure, which can lead to asymptotically faster
factorization algorithm for some matrices.
% However, assuming the ranks of the respective
% representations are bounded, the factorization complexity and the
% memory requirements are asymptotically lower for HSS than for BLR or HODLR.
\ignore{
\todo[inline]{[FH] The above sentence is true for some matrices and false
for some others. I think it is very dangerous to state things
like that or to try to justify why we do HSS and not something else.
So far no one has done any comparison besides a couple experiments
with BLR vs. HSS for 3D Helmholtz.}
}

Furthermore, thanks to the randomized HSS construction, our solver is
also fully structured (compared to partially structured approaches
where only part of the frontal matrix is compressed,
see~\cite{wang20113d}) and the larger frontal matrices are never
explicitly formed as dense matrices.

Achieving high performance on multi/many-core architectures can be
challenging but it has been demonstrated by many authors now that
dynamic scheduling of fine-grained tasks represented by a Directed
Acyclic Graph (DAG) can lead to good performance for a range of codes.
% It makes codes easier to write and maintain and also makes them more
% portable to other (for instance heterogeneous) architectures.
This approach was used successfully in the dense linear algebra
libraries PLASMA and MAGMA~\cite{agullo2009numerical} and more
recently it has become clear that it is also a convenient and
efficient strategy for sparse direct solvers. For instance,
in~\cite{lacoste2014taking} the PaStiX solver~\cite{henon2002pastix}
is modified to use two different generic DAG schedulers
(PaRSEC~\cite{bosilca2012dague} and
StarPU~\cite{augonnet2011starpu}). In~\cite{agullo2013multifrontal}
StarPU is used in a multifrontal QR solver.
% SuiteSparseQR~\cite{davis2011algorithm} uses Intel\tm{} TBB
% for task based parallel handling of the tree structure containing
% the frontal matrices.
In~\cite{kim2014parallel}, OpenMP tasks are submitted recursively for
work on different frontal matrices, while parallelism inside the
frontal matrices is also exploited with OpenMP tasks but with a manual
resolution of inter-task dependencies. The sparse Cholesky solver
HSL\_MA87~\cite{hogg2010design} uses a custom DAG scheduler
implemented in OpenMP. Just as sparse direct solvers, hierarchical
matrix algorithms also benefit from task parallelism:
Kriemann~\cite{kriemann2013} uses a DAG to schedule fine-grained tasks
to perform $\mathcal{H}$-matrix LU factorization. Our implementation
uses OpenMP task scheduling, but since most of the tasks are generated
recursively, a DAG is never explicitly constructed.

The main contribution of this work is the development of a robust and
efficient code for the solution of general sparse linear systems, with
a specific emphasis on systems from the discretization of PDEs.
% The algorithm was mostly taken
% from~\cite{xia2013randomized},~\cite{martinsson2011fast}
% and~\cite{xia2012superfast}.
Our work addresses various implementation issues, the most important
being the use of an adaptive HSS construction scheme
(Section~\ref{sec:HSS_comp_adaptive}), based on the randomized
sampling method~\cite{martinsson2011fast}.  Rather than assuming that
the maximum rank in the HSS submatrices is known a-priori, it is
computed in an adaptive way during the HSS compression. Other
implementation techniques such as fast extraction of elements from an
HSS structure (Section~\ref{sec:HSS_extraction}) are also
indispensable to make the code robust and usable as a black-box
solver.  The code achieves high performance and good scalability on a
range of modern multi/many-core architectures like Intel\tm{} Xeon and
Intel\tm{} Xeon Phi (MIC), due to runtime scheduled task parallelism,
using
OpenMP\footnote{\url{http://openmp.org/wp/openmp-specifications/}}.
The exclusive use of task parallelism avoids expensive inter-thread
synchronization and leads to a very scalable code.  This is the first
parallel algebraic sparse solver with fully structured HSS low-rank
approximation. The code is made publicly available with a BSD license
as part of a package called
STRUMPACK\footnote{\url{http://portal.nersc.gov/project/sparse/strumpack/}}
-- STRUctured Matrices PACKage. STRUMPACK also has a dense distributed
memory component, see~\cite{FHR}.

This work advances the field significantly on several fronts.
% compared to the prior work.
\begin{itemize}
\item Wang et al.~\cite{wang2014parallel} presented the first parallel
  multifrontal code with HSS embedding, called Hsolver. However, two
  shortcomings prevent it from being widely adopted: it is based on
  discretization on regular meshes and is only partially structured
  due to the hurdle of the extend-add of HSS update matrices (see
  Section~\ref{sec:MFHSS}).  Our new code, on the other hand, is a
  purely algebraic solver for general sparse linear systems and is
  fully structured, mitigating the HSS extend-add obstacle thanks to
  the randomized sampling technique (see
  Section~\ref{sec:skinny-extend-add}).
  % geometric distributed memory multifrontal code with
  % (non-randomized) HSS compression for finite difference
  % discretizations on regular meshes.  The code presented here is
  % shared memory parallel but it is an algebraic solver for general
  % sparse linear systems.
\item 
% The code was inspired by the StruMF code used in
  Napov and Li developed a purely algebraic sparse solver with HSS
  embedding~\cite{napov2013algebraic}, but it is only sequential and
  their experiments did not include the randomized sampling HSS
  compression.  Our new code is parallel and we show detailed results
  with randomized sampling.
\end{itemize}
In future work, building on the current paper and on the distributed
HSS code developed in~\cite{FHR}, we intend to develop a distributed
memory algebraic sparse solver with HSS compression.

The rest of the paper is outlined as follows.  Some required
background on HSS is briefly presented in
Section~\ref{sec:HSS}. First, in~\ref{sec:HSSformat}, the HSS
rank-structured format is described. Next, the fast randomized
sampling HSS construction~\cite{martinsson2011fast} and the ULV
decomposition~\cite{xia2012superfast} are discussed in
Sections~\ref{sec:HSSconstruction} and~\ref{sec:ULV} respectively.
Section~\ref{sec:MF} describes multifrontal LU decomposition. Then, in
Section~\ref{sec:MFHSS} we discuss how HSS matrices can be
incorporated into a multifrontal
solver. Section~\ref{sec:implementation} explains various aspects of
the actual implementation. In Section~\ref{sec:experiments} we present
experimental results that illustrate numerical and performance aspects
of the code. Finally, Section~\ref{sec:conclusions} has some
concluding remarks and an outlook to planned future work.

\section{HSS: Hierarchically Semi-Separable matrices}\label{sec:HSS}
This section briefly introduces hierarchically semi-separable (HSS)
matrices, mostly following the notation
from~\cite{martinsson2011fast}. HSS is a data-sparse matrix
representation which is part of the more general class of
$\mathcal{H}$-matrices and more specifically $\mathcal{H}^2$-matrices.

\subsection{Overview of the HSS matrix format}\label{sec:HSSformat}
The following notation is used: `$:$` is matlab-like notation for all
indices in the range, $^*$ denotes complex conjugation, $\#I_\tau$ is
the number of elements in index set $I_{\tau} = \{ i_1, i_2, \cdots,
i_n\}$ and $R_\tau = R(I_\tau,:)$ is the matrix consisting of only the
rows $I_\tau$ of matrix $R$.

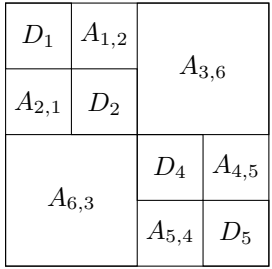
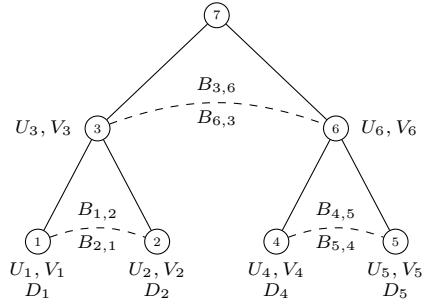
\begin{figure}
  \begin{subfigure}[b]{.48\textwidth}
    \begin{center}
      \begin{tikzpicture}[scale=3.5]
        \path[use as bounding box] (-.1,-.1) rectangle (1.1,1.1); % adjust to fit
        \draw (0,0) rectangle (1,1);
        \draw (0,0) rectangle (0.5,0.5);
        \draw (1,1) rectangle (0.5,0.5);
        \draw (0,1) rectangle (0.25,0.75);
        \draw (0.5,0.5) rectangle (0.75,0.25);
        \draw (0.25,0.75) rectangle (0.5,0.5);
        \draw (0.75,0.25) rectangle (1,0);
        \node at (0.125,0.875) {$D_1$};
        \node at (0.375,0.615) {$D_2$};
        \node at (0.625,0.375) {$D_4$};
        \node at (0.875,0.125) {$D_5$};
        \node at (0.125,0.615) {$A_{2,1}$};
        \node at (0.375,0.875) {$A_{1,2}$};
        \node at (0.625,0.125) {$A_{5,4}$};
        \node at (0.875,0.375) {$A_{4,5}$};
        \node at (0.25,0.25) {$A_{6,3}$};
        \node at (0.75,0.75) {$A_{3,6}$};
      \end{tikzpicture}
    \end{center}
    \caption{\footnotesize HSS partitioning of a square matrix.}
    \label{fig:HSSpartition}
  \end{subfigure}
  \begin{subfigure}[b]{.48\textwidth}
    \begin{tikzpicture}\scriptsize
      \node[cnode](7){7}
      child{node[cnode](3){3} child{node[cnode](1){1}} child{node[cnode](2){2}} }
      child{node[cnode](6){6} child{node[cnode](4){4}} child{node[cnode](5){5}} }  ;
      % Level-arcs and their labels
      \foreach \ln / \rn in {1/2,4/5,3/6}{
        \draw[dashed] (\ln) to [out=20,in=160] node[above]{\(B_{\ln,\rn}\)} node[below]{\(B_{\rn,\ln}\)} (\rn);
      }
      % Leaves labels
      \foreach \ln in {1,2,4,5}{ \node[below=.5em] at (\ln) {\(\begin{array}{c}U_{\ln},V_{\ln}\\D_{\ln}\end{array}\)}; }
      \foreach \ln in {3}{  \node[left=.2em] at (\ln) {\(\begin{array}{c}U_{\ln},V_{\ln}\end{array}\)}; }
      \foreach \ln in {6}{ \node[right=.2em] at (\ln) {\(\begin{array}{c}U_{\ln},V_{\ln}\end{array}\)}; }
    \end{tikzpicture}
    \caption{\footnotesize Tree corresponding to the HSS partition.}
    \label{fig:tree}
  \end{subfigure}
  \caption{\footnotesize Illustration of an HSS partitioning of a square
    matrix. Diagonal blocks are partitioned recursively. Figure (b)
    shows the tree, using postordering, corresponding to the
    partitioning in (a) and it illustrates the basis matrices stored
    in the nodes of the tree.}
  \label{fig:HSSillustration}
\end{figure}

Consider a square matrix $A \in \mathbb{C}^{N \times N}$ with an index
set $I_{A} = \{1,\dots,N \}$ associated with it. Let $\mathcal{T}$ be
a postordered binary tree, meaning that children in the tree are
numbered before their parent. Each node $\tau$ of the tree is
associated with a contiguous subset $t_\tau \subset \mathcal{I}$. For
two siblings in the tree, $\nu_1$ and $\nu_2$, children of $\tau$, it
holds that $t_{\nu_1} \cup t_{\nu_2} = t_\tau$ and $t_{\nu_1} \cap
t_{\nu_2} = \emptyset$. Furthermore,
$\cup_{\tau=\textrm{leaf}(\mathcal{T})} t_\tau =
t_{\textrm{root}(\mathcal{T})} = I_A$. The same tree $\mathcal{T}$ is
used for the rows and the columns of $A$ and only diagonal blocks are
partitioned. An example of the resulting matrix partitioning is given
in Figure~\ref{fig:HSSpartition} and the corresponding tree is shown
in Figure~\ref{fig:tree}.

% For the example in Figure~\ref{fig:HSSillustration} the index
% vectors are $I_7 = \{ 0:N \}$, $I_3 = \{0:N/2\}$, $I_6 = \{N/2:N
% \}$, $I_1 = \{0:N/4 \}$, $I_2 = \{N/4 : N/2\}$, $I_4 = \{ N/2 :
% 3N/4\}$ and $I_5 = \{ 3N/4 : N\}$. In the multifrontal solver, the
% HSS partition sizes do not have to be uniform and they are
% determined using graph partitioning routines, see
% Section~\ref{sec:MFHSS}.

The diagonal blocks of $A$, denoted $D_\tau$, are stored as dense
matrices in the leaves $\tau$ of the tree $\mathcal{T}$
\begin{equation}
  D_{\tau} = A(I_{\tau},I_{\tau}) \, .
\end{equation}
The off-diagonal blocks $A_{\nu_1,\nu_2} = A(I_{\nu_1},I_{\nu_2})$,
where $\nu_1$ and $\nu_2$ denote two siblings in the tree, are
factored (approximately) as
\begin{equation}
  A_{\nu_1,\nu_2} \approx U^{\mathrm{big}}_{\nu_1} B_{\nu_1,\nu_2} \left( V_{\nu_2}^{\mathrm{big}}\right)^*  \, .
\end{equation}
The matrices $U^{\mathrm{big}}_{\nu_1}$ and
$V_{\nu_2}^{\mathrm{big}}$, which form bases for the column and row
spaces of $A_{\nu_1,\nu_2}$, are typically tall and skinny, with
$U^{\mathrm{big}}_{\nu_1}$ having $\#I_{\nu_1}$ rows and $r^r_{\nu_1}$
(column-rank) columns, $V_{\nu_2}^{\mathrm{big}}$ has $\#I_{\nu_2}$
rows and $r^c_{\nu_2}$ (row-rank) columns and hence $B_{\nu_1,\nu_2}$
is $r^r_{\nu_1} \times r^c_{\nu_2}$. The HSS-rank $r$ of matrix $A$ is
defined as the maximum of $r^r_\tau$ and $r^c_\tau$ over all
off-diagonal blocks, where typically $r \ll N$. The matrices
$B_{\nu_1,\nu_2}$ and $B_{\nu_2,\nu_1}$ are stored in the parent of
$\nu_1$ and $\nu_2$. For a non-leaf node $\tau$ with children $\nu_1$
and $\nu_2$, the basis matrices $U^{\mathrm{big}}_\tau$ and
$V^{\mathrm{big}}_\tau$ are not stored directly since they can be
represented hierarchically as
\begin{equation}\label{eq:nestedUV}
  U^{\mathrm{big}}_{\tau} = \begin{bmatrix} U^{\mathrm{big}}_{\nu_1} & 0 \\ 0 & U^{\mathrm{big}}_{\nu_2} \end{bmatrix} U_\tau
  \quad \textrm{and} \quad
  V^{\mathrm{big}}_{\tau} = \begin{bmatrix} V^{\mathrm{big}}_{\nu_1} & 0 \\ 0 & V^{\mathrm{big}}_{\nu_2} \end{bmatrix} V_\tau \, .
\end{equation}
Note that for a leaf node $U^{\mathrm{big}}_\tau = U_\tau$ and
$V^{\mathrm{big}}_\tau = V_\tau$. Hence, every node $\tau$ with
children $\nu_1$ and $\nu_2$, except for the root node, keeps matrices
$U_\tau$ and $V_{\tau}$. The example from
Figure~\ref{fig:HSSillustration} can be written out explicitly as
\begin{equation}
  % A = \left[ \begin{array}{cc|cc}
  %   D_1 & U_1 B_{1,2} V_2^* & \multicolumn{2}{c}{\multirow{2}{*}{$\begin{bmatrix} U_1 & 0 \\ 0 & U_2 \end{bmatrix} U_3 B_{3,6} V_6^* \begin{bmatrix} V_4^* & 0 \\ 0 & V_5^* \end{bmatrix}$}} \\
  %   U_2 B_{2,1} V_1^* & D_2 &  \\ \hline
  %   \multicolumn{2}{c|}{\multirow{2}{*}{$\begin{bmatrix} U_4 & 0 \\ 0 & U_5 \end{bmatrix} U_6 B_{6,3} V_3^* \begin{bmatrix} V_1^* & 0 \\ 0 & V_2^* \end{bmatrix}$}} 
  %   & D_4 & U_4 B_{4,5} V_5^* \\
  %     & & U_5 B_{5,4} V_4^* & D_5
  % \end{array}\right] \, .
  A = \begin{bmatrix}
    D_1 & U_1 B_{1,2} V_2^* & \multicolumn{2}{c}{\multirow{2}{*}{$\begin{bmatrix} U_1 & 0 \\ 0 & U_2 \end{bmatrix} U_3 B_{3,6} V_6^* \begin{bmatrix} V_4^* & 0 \\ 0 & V_5^* \end{bmatrix}$}} \\
    U_2 B_{2,1} V_1^* & D_2 &  \\
    \multicolumn{2}{c}{\multirow{2}{*}{$\begin{bmatrix} U_4 & 0 \\ 0 & U_5 \end{bmatrix} U_6 B_{6,3} V_3^* \begin{bmatrix} V_1^* & 0 \\ 0 & V_2^* \end{bmatrix}$}} 
    & D_4 & U_4 B_{4,5} V_5^* \\
      & & U_5 B_{5,4} V_4^* & D_5
    \end{bmatrix} \, .
    \label{eq:HSSexamplematrix}
\end{equation}

The storage requirements for an HSS matrix are
$\mathcal{O}(rN)$. Construction of the HSS generators will be
discussed in the next section. Once an HSS representation of a matrix
is available, it can be used to perform matrix-vector multiplication
in $\mathcal{O}(rN)$ operations compared to $\mathcal{O}(N^2)$ for
classical dense matrix-vector multiplication,
see~\cite{martinsson2011fast,FHR}.

\subsection{Fast HSS construction through randomized sampling}\label{sec:HSSconstruction}
In~\cite{martinsson2011fast} Martinsson presents a randomized sampling
algorithm for the efficient construction of an HSS representation of a
matrix $A$. Note that the same technique was also used by Xia et
al. in~\cite{xia2012superfast,xia2013randomized} for HSS compression
in a multifrontal solver. The main advantage of this approach is that
it does not need explicit access to all elements of $A$, but only
needs a fast matrix-vector routine and selected elements from $A$. The
matrix $A$ never needs to be formed explicitly as a dense matrix and
this allows to save memory. The overall complexity of the algorithm is
$\mathcal{O}(N r^2)$, with $r$ the HSS-rank of $A$, provided that a
fast ($\mathcal{O}(N)$) matrix-vector product is available. By
comparison, other approaches based on direct low-rank compression of
matrix off-diagonal blocks typically require $\mathcal{O}(N^2 r)$
operations. This section briefly presents the randomized compression
algorithm, for a more in depth discussion
see~\cite{FHR,martinsson2011fast}.

Suppose the HSS-rank $r$ is known a priori and $R \in \mathbb{C}^{N
  \times d}$ is a tall and skinny random matrix with $d = r + p$
columns where $p$ is a small oversampling parameter.
% , for instance $p=10$.
Let $S^r = A R$ and $S^c = A^* R$ be samples for the row (superscript
$^r$) and column bases (superscript $^c$) of $A$
respectively. Algorithm~\ref{algo:compress} with $R^r \equiv R^c
\equiv R$ computes the HSS representation of $A$ using the information
available in the samples $S^r$ and $S^c$ by hierarchically compressing
(using interpolative decompositions, see below) the off-diagonal
blocks of $A$, starting from the leaves.

Let $D_\tau$ for a non-leaf node $\tau$ with children $\nu_1$ and
$\nu_2$  be defined as
\begin{equation}
  D_\tau = \begin{bmatrix} D_{\nu_1} & A_{\nu_1,\nu_2} \\ A_{\nu_2,\nu_1} & D_{\nu_2} \end{bmatrix} \, .
\end{equation}
If $\{\tau_1,\tau_2,\dots,\tau_q\}$ are all the nodes on level $\ell$
of the HSS tree, then 
\begin{equation}
  D^{(\ell)} = \mathrm{diag}(D_{\tau_1}, D_{\tau_2},\dots,D_{\tau_q})
\end{equation}
is an $N \times N$ block diagonal matrix. The main idea of the
randomized sampling Algorithm~\ref{algo:compress} is to construct a
sample matrix $S^{(\ell)}$ for each level of the tree as
\begin{equation}
  S^{(\ell)} = \left( A - D^{(\ell)} \right) R = S^r - D^{(\ell)} R \, .
\end{equation}
This sample matrix $S^{(\ell)}$ captures the action of a product of
the block off-diagonal part of $A$ with a set of random vectors
$R$. It is exactly this block off-diagonal part that needs to be
compressed using low-rank approximation.
% $\ell = L$ is the finest level.

Another crucial component of the randomized sampling algorithm is the
interpolative decomposition (ID)~\cite{cheng2005compression}. The ID
computes a factorization of a rank-$k$ matrix $Y \in \mathbb{C}^{m
  \times n}$ by expressing $Y$ as a linear combination of a set $J$ of
$k$ selected columns of $Y$:
\begin{equation}
  \left[ X, J \right] = \mathrm{ID}(Y), \quad \textrm{such that} \quad 
  Y = Y(:,J) X, \,\, Y(:,J) \in \mathbb{C}^{m \times k}, \,\, X \in \mathbb{C}^{k \times n} \, ,
\end{equation}
or it can be modified to take a compression tolerance $\varepsilon$,
such that
\begin{equation}
  \left[ X, J \right] = \mathrm{ID}(Y, \varepsilon), \,\, \textrm{s.t.} \,\, 
  Y = Y(:,J) X + \mathcal{O}(\varepsilon), \,\, Y(:,J) \in \mathbb{C}^{m \times k'}, \,\,
  X \in \mathbb{C}^{k' \times n} \, ,
\end{equation}
where $k' \leq k$ is the numerical rank. The ID can be computed from a
rank-revealing or column pivoted QR
decomposition~\cite{chan1987rank,quintana1998blas}
\begin{equation}
  Y \Pi = Q \begin{bmatrix} R_1 & R_2 \end{bmatrix} \, ,
\end{equation}
where $R_1$ is upper-triangular, followed by a triangular solve such
that
\begin{equation}
  Y = \left (QR_1 \right) \left( \begin{bmatrix} I & R^{-1}_1 R_2 \end{bmatrix} \Pi^{-1} \right) \equiv Y(:,J) X \, .
\end{equation}

\begin{algorithm}[!ht]
  \DontPrintSemicolon
  \SetAlgoLined
  \SetKwProg{Fn}{Function}{}{}
  \SetKwFunction{compress}{HSSCompress}

  \Fn{$A_{\mathrm{hss}} = $ \compress{$R^r$, $R^c$, $S^r$, $S^c$, $\varepsilon$, $\tau = \mathrm{root}(A_{\mathrm{hss}})$}}{
    \KwData{%$d = r + p$ with $r$ an upper bound for the rank of $A \in \mathbb{R}^{N \times N}$\newline
      $S^r = AR^r$ and $S^c = A^* R^c$ with $\{ S^r, S^c, R^r, R^c \} \in \mathbb{R}^{N \times d}$, $d \geq r_{\max} + p$
      %A tree on the index vector $[1, \dots, N]$\newline
      %Index set $I_\tau$ for each HSS node $\tau$ of size $\#I_\tau = b_\tau$
    }
    \KwResult{$A_{\mathrm{hss}}$: $D_\tau$ (leaves), $B_{\nu_1,\nu_2}$, $B_{\nu_2, \nu_1}$ (non-leaves), $U_\tau$, $V_\tau$
      (all except root).}
    \BlankLine
    \lForEach{$\nu \in \mathrm{child}(\tau)$}{\compress{$R^r$, $R^c$, $S^r$, $S^c$, $\nu$}}
    \uIf{$\mathrm{child}(\tau) \equiv \emptyset$}{
      $D_\tau = A(I_\tau, I_\tau)$\;
      $S^r_{\tau} = S^r(I_\tau,:) - D_\tau R^r(I_\tau,:)$ \hfill $S^c_{\tau} = S^c(I_\tau,:) - D_\tau^* R^c(I_\tau,:)$ \label{algoline:updateSleaf} \;
    }
    \uElse(\tcp*[h]{$\nu_1$ and $\nu_2$ are the children of node $\tau$}){
      $B_{\nu_1,\nu_2} = A(I^r_{\nu_1}, I^c_{\nu_2})$ \hfill $B_{\nu_2,\nu_1} = A(I^r_{\nu_2}, I^c_{\nu_1})$ \;
      $S^r_{\tau} = \begin{bmatrix} S^r_{\nu_1} - B_{\nu_1,\nu_2} R^r_{\nu_2}\\
        S^r_{\nu_2} - B_{\nu_2,\nu_1} R^r_{\nu_1} \end{bmatrix}$ \hfill
      $S^c_{\tau} = \begin{bmatrix} S^c_{\nu_1} - B^*_{\nu_2,\nu_1} R^c_{\nu_2}\\
        S^c_{\nu_2} - B^*_{\nu_1,\nu_2} R^c_{\nu_1} \end{bmatrix}$ \label{algoline:updateSnonleaf}\;
    }
    $\begin{bmatrix} U_\tau^*, J^r_\tau \end{bmatrix} = \mathrm{ID}((S^r_{\tau})^*, \varepsilon )$ \hfill
    \quad\,\,$\begin{bmatrix} V_\tau^*, J^c_\tau \end{bmatrix} = \mathrm{ID}((S^c_{\tau})^*, \varepsilon)$ \label{algoline:id}\;
    $S^r_\tau \gets S^r_{\tau}(J^r_\tau,:)$ \hfill
    \quad\,\,$S^c_\tau \gets S^c_{\tau}(J^c_\tau,:)$ \label{algoline:updateSID}\;
    \uIf{$\mathrm{child}(\tau) \equiv \emptyset$}{
      $R^r_\tau = V_\tau^* R^r(I_\tau,:)$ \hfill
      $R^c_\tau = U_\tau^* R^c(I_\tau,:)$ \label{algoline:updateRleaf} \;
      $I^r_\tau = I_{\tau}(J^r_\tau)$ \hfill
      $I^c_\tau = I_{\tau}(J^c_\tau)$ \;
    }
    \uElse{
      $R^r_\tau = V_\tau^* \begin{bmatrix} R^r_{\nu_1} \\ R^r_{\nu_2}\end{bmatrix}$ \hfill 
      $R^c_\tau = U_\tau^* \begin{bmatrix} R^c_{\nu_1} \\ R^c_{\nu_2}\end{bmatrix}$ \label{algoline:updateRnonleaf} \;
      $I^r_\tau = [I^r_{\nu_1} \,\, I^r_{\nu_2} ](J^r_\tau)$ \hfill
      $I^c_\tau = [I^c_{\nu_1} \,\, I^c_{\nu_2} ](J^c_\tau)$ \;
    }
  }
  \caption{\footnotesize Computing the HSS factorization of a nonsymmetric
    matrix. }
  \label{algo:compress}
\end{algorithm}

A consequence of using the ID in Algorithm~\ref{algo:compress} is that
$B_{\nu_1,\nu_2} = A(I^r_{\nu_1}, I^c_{\nu_2})$ is a submatrix of the
original matrix $A$. Furthermore, it also leads to a special structure
for the $U_\tau$ and $V_\tau$ generators:
\begin{equation}
  U_\tau = \Pi^r_\tau \begin{bmatrix} I \\ E^r_\tau \end{bmatrix} \quad \textrm{and} \quad 
  V_\tau = \Pi^c_\tau \begin{bmatrix} I \\ E^c_\tau \end{bmatrix} \, ,
  \label{eq:UVstructure}
\end{equation}
referred to as interpolative bases, which can be exploited in the
computations. Note that these interpolative bases are not
orthonormal. Although creating orthonormal bases might slightly
improve stability, the interpolative structure improves performance of
the compression algorithm and the ULV decomposition, see
Section~\ref{sec:ULV}.

\subsubsection{Adaptive scheme to determine the HSS-rank}\label{sec:HSS_comp_adaptive}
In practice however, the HSS-rank of the matrix is not known in
advance. In this case, Algorithm~\ref{algo:compress} can be called
repeatedly while increasing the number of columns of $R$, $S^r$ and
$S^c$. As long as $d < r+p$, the ID in line~\ref{algoline:id} will
fail. Suppose the ID fails at node $\tau$, i.e., the required accuracy
$\varepsilon$ is not reached, but the descendants of node $\tau$ are
successfully compressed. In that case, during the next iteration of
Algorithm~\ref{algo:compress} with $d \leftarrow d + \Delta d$, it is
not necessary to redo the compression (ID) or the extraction of $D$
and $B$ for the descendants of node $\tau$. However, those descendants
do have to update the $\Delta d$ new columns in $R^{r/c}$
(lines~\ref{algoline:updateRleaf} and~\ref{algoline:updateRnonleaf})
and $S^{r/c}$ (lines~\ref{algoline:updateSleaf},
\ref{algoline:updateSnonleaf}
and~\ref{algoline:updateSID}). In~\cite{FHR}, this adaptive rank
scheme is presented in more detail.

\ignore{
\subsubsection{Improving the samples by power iteration}\label{sec:power_iteration}
When the singular values of $A$ do not decay fast or $A$ is very
large, the samples $S^r$ and $S^c$ can be poor approximations of the
column and row spaces of $A$. In that case, sampling can be done using
$B = (A A^*)^q A$ instead of $A$, with $q$ a small integer. Note that
$B$ has the same singular vectors as $A$ but it's singular values
$\sigma_j(B) = \sigma_j(A)^{2q+1}$, $j=1,2,\dots$ decay much
faster. Let
\begin{equation}
  S^r = \left( A A^* \right)^q A R = %A A^* \left( A A^* \right)^{q-1} A R = 
  A R^r \quad \textrm{and} \quad
  S^c = \left( A^* A \right)^q A^* R = % A^* A \left( A^* A \right)^{q-1} A^* R =
  A^* R^c , \label{eq:poweritsample}
\end{equation}
then Algorithm~\ref{algo:compress} can be called with $R^r = A^*
\left( A A^* \right)^{q-1} A R$, $R^c = A \left( A^* A \right)^{q-1}
A^* R$. Typically, $q=1$ or $2$ should be sufficient, so when a fast
matrix-vector multiplication is available for $A$, the additional
computational cost is not too bad. However, it might also be necessary
to reorthogonalize the random vectors between each application of $A$
and $A^*$. See~\cite{halko2011finding} for more info.
% We add this power iteration as an option in the code.  
% theorem about accuracy/probability, include $q$ in the theorem
}

\subsubsection{Implementation issues}
The random matrices $R^r$ and $R^c$ are filled element by element
using a pseudo-random number generator. Our implementation offers the
\verb+minstd_rand+ and \verb+mt19937+ generators from the
C\texttt{++}11 standard while the distribution can be either uniform
over $[0,1)$ or standard normal (Gaussian) $\mathcal{N}(0,1)$. By
default the linear congruential engine\footnote{This choice is
  motivated further in section~\ref{sec:MFHSS}.} \verb+minstd_rand+ is
selected in combination with the Gaussian distribution.

\ignore{ Alternatively, so-called \emph{structured} random matrices,
  for instance based on the subsampled randomized Fourier or Hadamard
  transforms~\cite{halko2011finding}, might be considered for
  $R$. Such matrices allow faster $AR$ evaluation when $A$ is a dense
  matrix.}

The rank-revealing QR factorization, used in the ID, could be replaced
by a \emph{strong} rank-revealing QR
factorization~\cite{gu1996efficient}, with possibly greater accuracy
and smaller HSS-rank. This is left as future work. Two interesting
alternative approaches to the randomized compression routine discussed
in this section should be mentioned, namely adaptive cross
approximation~\cite{bebendorf2000approximation} and a matrix-free
approach presented in~\cite{lin2011fast}.

\subsection{ULV-like factorization and solve}\label{sec:ULV}
Solving a linear system with an HSS matrix can be done by first
computing a so-called ULV decomposition~\cite{chandrasekaran2006fast},
where $U$ and $V^*$ are unitary matrices and $L$ is lower
triangular. However, in~\cite{xia2013randomized}
and~\cite{xia2012superfast}, the ULV decomposition is modified to take
advantage of the special structure of the $U_\tau$ and $V_\tau$
generators, see~\eqref{eq:UVstructure}. The resulting algorithm is
referred to as ULV-like since it is no longer based on unitary
transformations.

In the first step of a ULV factorization, zeros are introduced in the
HSS block rows. This step can be done using for instance a full QL
factorization
\begin{equation}
  U_i = \Omega_{\tau} \begin{bmatrix} 0 \\ \tilde{U}_i \end{bmatrix}, \quad
  \Omega_{\tau}^* U_i = \begin{bmatrix} 0 \\ \tilde{U}_i \end{bmatrix} \, .
\end{equation}
However, thanks to the special structure of $U_\tau$, a multiplication
from the left with a carefully chosen $\Omega_\tau$ is much cheaper
and has a similar effect
\begin{equation}
  \Omega_\tau = \begin{bmatrix} -E^r_\tau & I \\ I & 0 \end{bmatrix} {\Pi_\tau^r}^T \quad \rightarrow \quad 
  \Omega_\tau U_\tau = \Omega_{\tau} \Pi^r_\tau \begin{bmatrix} I \\ E^r_\tau\end{bmatrix} = 
%  \begin{bmatrix} -E^r_\tau & I \\ I & 0 \end{bmatrix} {\Pi_\tau^r}^T \Pi^r_\tau \begin{bmatrix} I \\ E^r_\tau\end{bmatrix} =
  \begin{bmatrix} 0 \\ I \end{bmatrix} \, .
\end{equation}
We refer to~\cite{FHR} for a detailed description of the ULV
factorization and the corresponding solve.

\section{Multifrontal sparse LU factorization}\label{sec:MF}
This section briefly recalls the main ingredients of the multifrontal
method for the LU factorization of general invertible sparse
matrices. For a more detailed discussion of multifrontal methods,
see~\cite{duff1983multifrontal,liu1992multifrontal}.  The method casts
the factorization of a sparse matrix into a series of partial
factorizations of many smaller dense matrices and Schur complement
updates.
% We are to compute a factorization of a given matrix $A$, $A=LU$ if the
% matrix is non-symmetric, or $A=LDL^T$ if the matrix is symmetric.

\subsection{Matrix reordering}
As a preprocessing step, $A$ is first scaled and permuted for
numerical stability: $A \gets D_r A D_c Q_c$, where $D_r$ and $D_c$
are diagonal matrices that scale the rows and columns of $A$ and $Q_c$
is a column permutation that places large entries on the diagonal. We
use the MC64 code by Duff and Koster~\cite{duff1999design} to perform
the scaling and column permutation. Popular alternative scaling
algorithms can be found
in~\cite{ruiz2001scaling,amestoy2008parallel,curtis1972automatic}.
After that, a fill-reducing permutation $A \gets P A P^T$ is applied
in order to reduce the number of nonzero elements in the LU
factors. Permutation matrix $P$ is computed using nested dissection
applied to the adjacency graph of $A+A^T$, using one of the graph
partitioning tools SCOTCH~\cite{pellegrini1996scotch} or
METIS~\cite{karypis1998fast}. Instead of nested dissection, other
heuristics like AMD~\cite{amestoy1996amd} can be used.

The multifrontal method relies on a structure called the
\emph{elimination tree}. The elimination tree serves as a task and
data-dependency graph for both the factorization and the solution
process. A few equivalent definitions of the elimination tree are
available. We use the following, and we recommend the survey by
Liu~\cite{liu1992multifrontal} for more detail on the method and the
survey by L'Excellent for more detail about implementation issues like
parallelism, memory usage, numerical aspects
etc.~\cite{lexcellent2012multifrontal}.
\begin{mydef}
  Assume $A=LU$, where $A$ is an $N \times N$ sparse, structurally
  symmetric matrix. The elimination tree of $A$ is a tree with $N$
  nodes, where the $i$-th node corresponds to the $i$-th column of $L$
  and with the parent relations defined by: $\mathrm{parent}(j) =
  \min\{i: i>j \;\mathrm{and}\; \ell_{ij}\neq 0\}$, for $j=1,\ldots,
  N-1$.
\end{mydef}

% Without loss of generality, we assume that $A$ is irreducible.
In practice, nodes are amalgamated: nodes that represent columns and
rows of the factors with similar structures are grouped together in a
single node. For instance when using nested dissection reordering, all
vertices from the same graph separator can be grouped in one
elimination tree node. In the end, each node corresponds to a square
dense matrix, referred to as a \emph{frontal matrix}, with the
following $2 \times 2$ block structure:
\begin{equation}\label{eqn:frontal}
  \mathcal{F}_i = \begin{bmatrix}F_{11} & F_{12}\\ F_{21} & F_{22}\end{bmatrix} \, .
\end{equation}

\subsection{Numerical factorization}
Multifrontal factorization of the matrix consists in a bottom-up
traversal of the tree, following a topological order (a node is
processed before its parent). Processing a node means first forming
(or \emph{assembling}) the frontal matrix followed by elimination of
the fully-summed variables in the $F_{11}$ block and finally a Schur
complement update step. The frontal matrix $\mathcal{F}_i$ is formed
by summing the rows and columns of $A$ corresponding to the variables
in the $F_{11}$, $F_{21}$ and $F_{12}$ blocks, with the temporary
data -- the extended update matrices $\bar{\mathcal{U}}_\nu$ -- that have
been produced by the children of $i$ after their elimination step,
i.e.,
\begin{equation}\label{eqn:frontal_1}
  \mathcal{F}_i = {A}_i + \sum_{\nu \, \in \, \mathrm{child}(i)} \bar{\mathcal{U}}_\nu =
  \begin{bmatrix} 
    A(I^{\mathrm{sep}}_i,I^{\mathrm{sep}}_i) & A(I^{\mathrm{sep}}_i,I^{\mathrm{upd}}_i) \\ 
    A(I^{\mathrm{upd}}_i,I^{\mathrm{sep}}_i) & 
  \end{bmatrix} + \bar{\mathcal{U}}_{\nu_1} + \bar{\mathcal{U}}_{\nu_2} + \dots \, ,
\end{equation}
where $I_i = \{ I_i^{\mathrm{sep}}, I_i^{\mathrm{upd}} \}$ is the set
of row and column indices of $\mathcal{F}_i$ w.r.t. the global matrix
$A$, after reordering. Eliminating the fully-summed variables in the
$F_{11}$ block is done through a partial factorization of
$\mathcal{F}_i$, typically via a standard dense matrix factorization
of the $F_{11}$ block.
% , which produces the corresponding rows and columns of the factors
% stored in $F_{11}$, $F_{21}$ and $F_{12}$.
Next, the Schur complement (contribution block or update matrix) is
computed as $\mathcal{U}_i= F_{22}-F_{21} F_{11}^{-1} F_{12}$ and
stored in temporary memory.
% ; it will be used to form the frontal matrix associated with the
% parent node.  Therefore, when a node is activated, it ``consumes''
% the contribution blocks of its children.
In contrast to the elimination step which uses straightforward dense
matrix operations (high performance LAPACK/BLAS3 codes), the assembly
step~\eqref{eqn:frontal_1} requires index manipulation and indirect
addressing while summing up $\mathcal{U}_k$. For example, if two
children's update matrices $\mathcal{U}_k = \begin{bmatrix} a_k & b_k
  \\ c_k & d_k \end{bmatrix}, k=\nu_1,\nu_2$, have subscript sets
$I^{\mathrm{upd}}_1=\{1,2\}$ and $I^{\mathrm{upd}}_2=\{1,3\}$,
respectively, then those update matrices can only be added after
aligning the index sets of the two matrices by padding with zero
entries
\begin{equation} \label{eqn:extend-add}
  \mathcal{U}_1 \extadd \mathcal{U}_2 = \bar{\mathcal{U}}_1 + \bar{\mathcal{U}}_2 = 
  \begin{bmatrix}
    a_{1} & b_{1} & 0\\
    c_{1} & d_{1} & 0\\
    0 & 0 & 0
  \end{bmatrix} +
  \begin{bmatrix}
    a_{2} & 0 & b_{2}\\
    0 & 0 & 0\\
    c_{2} & 0 & d_{2}
  \end{bmatrix} =
  \begin{bmatrix}
    a_{1}+a_{2} & b_{1} & b_{2}\\
    c_{1} & d_{1} & 0\\
    c_{2} & 0 & d_{2}%
  \end{bmatrix}.
\end{equation}
This summation operation is called \textit{extend-add}, denoted by
$\extadd$. The relationship between frontal matrices and update
matrices can be revealed by $\mathcal{F}_{i} = A_{i} \extadd
\mathcal{U}_{\nu_1} \extadd \mathcal{U}_{\nu_2} \extadd \cdots \extadd
\mathcal{U}_{\nu_q}$, where nodes $\nu_{1},\nu_{2},\ldots,\nu_{q}$ are
the children of $i$.

Each partial factorization might involve pivoting within the frontal
matrix. It can also happen that no suitable pivot can be found during
a step of partial factorization. In this situation, the corresponding
row and column remain unfactored and are sent to the parent node. This
strategy is used for instance in the MUMPS~\cite{amestoy2001fully}
code. Currently our code does not perform any such delayed pivoting,
but instead relies on static pivoting (using MC64) and partial
pivoting during the LU decomposition of the $F_{11}$ blocks.

\ignore{ %just remove this??
In the multifrontal factorization, the active memory, i.e., the memory
in use at a given step in the factorization, consists of the frontal
matrix being processed and a set of contribution blocks that are
temporarily stored and will be consumed at a later step.  The
multifrontal method lends itself very naturally to parallelism since
multiple processes can be employed to treat one, large enough, frontal
matrix or to process concurrently frontal matrices belonging to
separate subtrees.  These two sources of parallelism are commonly
referred to as node and tree parallelism, respectively, and their
correct exploitation is the key to achieving high performance on
parallel architectures.
}

%\subsection{Solution and complexity}
\subsection{Solution}
Once the factors are computed, the solution $x$ of $Ax = b$ is
computed in two steps: forward solution by doing a triangular solution
with the $L$ factor and backward substitution by doing a triangular
solution with the $U$ factor. The forward solution step is a bottom-up
topological traversal of the elimination tree, while the backward
substitution is a top-down traversal.

\section{Multifrontal solver with HSS frontal matrices}\label{sec:MFHSS}
This section explains how a multifrontal solver, see
Section~\ref{sec:MF}, can be used in combination with the HSS
data-structures and algorithms from Section~\ref{sec:HSS} to improve
the computational complexity and storage requirements. This section
closely follows~\cite{xia2013randomized}.

\subsection{Selection of HSS frontal matrices}
Note that the largest frontal matrices, those that determine the
computational complexity of the solver, typically correspond to nodes
closer to the root of the elimination tree. Let the top of the tree,
i.e., the root node, be at level $\ell=0$ of the tree. Then, define a
switch-level $\ell_s$ such that the frontal matrices at levels $\ell
\geq \ell_s$ of the elimination tree are stored as regular dense
matrices whereas those at levels $\ell < \ell_s$ are compressed using
the HSS format. According to the analysis in~\cite{xia2013randomized},
$\ell_s$ should be chosen such that the factorization cost above and
below the switch-level are equal. However, this rule is not very
practical and experiments show that performance depends crucially on
the choice of $\ell_s$.

\subsection{Separator reordering}\label{sec:sep_reorder}
Apart from the scaling and permutation of $A$ for stability, and
nested dissection reordering to reduce fill-in, an additional
reordering is applied to the index set of each separator. This
reordering is needed to obtain favorable HSS rank structure in the
corresponding frontal matrices. It is computed by recursively
bisecting the graph of the separator in subgraphs of size
approximately $b$ (defaults to $b=128$), using a graph partitioning
tool (SCOTCH or METIS). Each partition then corresponds to a leaf in
the HSS tree of the corresponding frontal matrix. However, since a
separator graph can be disconnected, it is enriched with length-two
connections from the connectivity graph before it is passed to the
partitioner, see also the discussion
in~\cite{napov2013algebraic}. Note that other reorderings can be used
instead of nested dissection. The influence of the reordering on the
ranks of off-diagonal blocks is studied
in~\cite{weisbecker2013improving}.

\subsection{Skinny extend-add}\label{sec:skinny-extend-add}
From here on, we assume that a binary elimination tree is used. The
steps followed for each HSS frontal matrix $\mathcal{F}_i$ are as
follows. First, a random matrix $R_i \in \mathbb{C}^{\#I_i \times
  d_i}$ is constructed. If the children $\nu_1$ and $\nu_2$ of $i$ are
also HSS, then $R_i$ is constructed as follows
\begin{equation}\label{eq:Rmerge}
  R_i(r,c) = 
  \begin{cases}
    R_{\nu_1}(r,c) \equiv R_{\nu_2}(r,c) & \textrm{if } c < \min(d_{\nu_1},d_{\nu_2}), \, I_i(r) \in I^{\mathrm{upd}}_{\nu_1} \textrm{ and } I_i(r) \in I^{\mathrm{upd}}_{\nu_2}, \\
    R_{\nu_1}(r,c) & \textrm{if } c < d_{\nu_1} \textrm{ and } I_i(r) \in I^{\mathrm{upd}}_{\nu_1}, \\
    R_{\nu_2}(r,c) & \textrm{if } c < d_{\nu_2} \textrm{ and } I_i(r) \in I^{\mathrm{upd}}_{\nu_2}, \\
    \textrm{random}(r,c) & \textrm{otherwise}
  \end{cases} \, .
\end{equation}
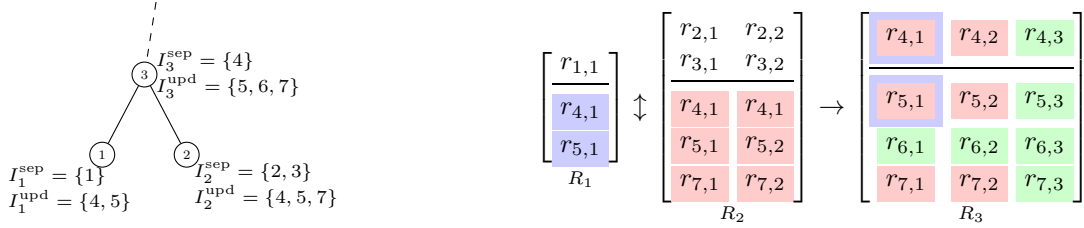
\begin{figure}
  % \begin{center}
  %   \includegraphics[width=0.8\textwidth]{figures/ext-add}
  % \end{center}
  \begin{subfigure}{.34\textwidth}
    \pgfdeclarelayer{background}
    \pgfdeclarelayer{foreground}
    \pgfsetlayers{background,main,foreground}
    \begin{center}
      \begin{tikzpicture}\scriptsize
        \node[](4){} 
        child[dashed]{node[solid,cnode,left](3){3} child[solid]{node[solid,cnode](1){1}} child[solid]{node[solid,cnode](2){2} }};
        \node[below=.2em,text width=2.5cm] at (1) {$I_1^{\textrm{sep}} = \{1\}$ \\ $I_1^{\textrm{upd}} = \{4,5\}$};
        \node[below right,text width=2.5cm] at (2) {$I_2^{\textrm{sep}} = \{2,3\}$ \\ $I_2^{\textrm{upd}} = \{4,5,7\}$};
        \node[right=.2em,text width=2.5cm] at (3) {$I_3^{\textrm{sep}} = \{4\}$ \\ $I_3^{\textrm{upd}} = \{5,6,7\}$};
      \end{tikzpicture}
    \end{center}
  \end{subfigure}
  \begin{subfigure}{.64\textwidth}
    \begin{center}
      \begin{equation*}
        \underset{R_1}{\begin{bmatrix}
            r_{1,1} \\
            \cmidrule(lr){1-1}
            \colorbox{blue!20}{$r_{4,1}$} \\
            \colorbox{blue!20}{$r_{5,1}$}
          \end{bmatrix}}
        \updownarrow
        \underset{R_2}{\begin{bmatrix}
            r_{2,1} & \!\!\!\!r_{2,2} \\
            r_{3,1} & \!\!\!\!r_{3,2} \\
            \cmidrule(lr){1-2}
            \colorbox{red!20}{$r_{4,1}$} & \!\!\!\!\colorbox{red!20}{$r_{4,1}$} \\
            \colorbox{red!20}{$r_{5,1}$} & \!\!\!\!\colorbox{red!20}{$r_{5,2}$} \\
            \colorbox{red!20}{$r_{7,1}$} & \!\!\!\!\colorbox{red!20}{$r_{7,2}$} \\
            %\colorbox{red!20}{$r_{8,1}$} & \!\!\!\!\colorbox{red!20}{$r_{8,2}$}
          \end{bmatrix}}
        \rightarrow
        \underset{R_3}{\begin{bmatrix}
            \colorbox{blue!20}{\colorbox{red!20}{$r_{4,1}$}} & \!\!\!\!\colorbox{red!20}{$r_{4,2}$} & \!\!\!\!\colorbox{green!20}{$r_{4,3}$} \\
            \cmidrule(lr){1-3}
            \colorbox{blue!20}{\colorbox{red!20}{$r_{5,1}$}} & \!\!\!\!\colorbox{red!20}{$r_{5,2}$} & \!\!\!\!\colorbox{green!20}{$r_{5,3}$} \\
            \colorbox{green!20}{$r_{6,1}$} & \!\!\!\!\colorbox{green!20}{$r_{6,2}$} & \!\!\!\!\colorbox{green!20}{$r_{6,3}$} \\
            \colorbox{red!20}{$r_{7,1}$} & \!\!\!\!\colorbox{red!20}{$r_{7,2}$} & \!\!\!\!\colorbox{green!20}{$r_{7,3}$} \\
            %\colorbox{red!20}{$r_{8,1}$} & \!\!\!\!\colorbox{red!20}{$r_{8,2}$} & \!\!\!\!\colorbox{green!20}{$r_{8,3}$} \\
          \end{bmatrix}}
      \end{equation*}   
    \end{center}
  \end{subfigure}
  \caption{\footnotesize Illustration of the extend-merge procedure for the
    random vectors. Node 3 needs 3 random vectors. It can get elements
    $r_{4,1}$ and $r_{5,1}$ from either child 1 or 2. Elements
    $r_{7,1}$, $r_{7,2}$, $r_{4,2}$ and $r_{5,2}$ are copied from
    child 2. Elements $r_{6,1}$ and $r_{6,2}$ in $R_3$ are generated
    with the properly seeded pseudo random number generator. When the
    adaptive HSS compression scheme decides that a third column has to
    be added to $R_3$, those elements are also generated. }
  \label{fig:extend-add}
\end{figure}
The random matrices of the children are merged in the parent $R_i$ and
any elements not present in any of the children's $R$ are
generated. This extend-merge procedure is illustrated in
Figure~\ref{fig:extend-add}. If node $i$ has no (HSS) children, $R_i$
is generated. However, it is important that corresponding ``random''
entries in $R_{\nu_1}$ and $R_{\nu_2}$ are equal, since that allows
efficient evaluation of $S^r_i = \mathcal{F}_{i} R_i$ (similarly for
$\mathcal{F}^*_{i} R_i$) based on
\begin{equation}\label{eq:FtimesR}
  \mathcal{F}_{i} R_i = \left( A_{i} \extadd \mathcal{U}_{\nu_1} \extadd \mathcal{U}_{\nu_2} \right) R_i 
  = \left( A_{i} R_i \right) \skinnyextadd \left( \mathcal{U}_{\nu_1} R_i(I^{\mathrm{upd}}_{\nu_1},:) \right) 
  \skinnyextadd \left( \mathcal{U}_{\nu_2} R_i(I^{\mathrm{upd}}_{\nu_2},:) \right) \,,
\end{equation}
where $R_i(I^{\mathrm{upd}}_{\nu_1},:)$ denotes the subset of rows of
$R_i$ which are also in $I^{\mathrm{upd}}_{\nu_1}$ and $\skinnyextadd$
denotes an extend-add operation where the extend is only done for the
rows, not the columns. By the construction of $R_i$~\eqref{eq:Rmerge},
the first $d_{\nu_1}$ columns of $R_i(I^{\mathrm{upd}}_{\nu_1},:)$ are
already available at node $\nu_{1}$, which is convenient for the
evaluation of $\mathcal{U}_{\nu_1} R_i(I^{\mathrm{upd}}_{\nu_1},:)$.
% Figure~\ref{fig:extend-add} illustrates the extend-merge operation for
% the random vectors $R$ and the extend-add for the samples $S_i =
% \mathcal{F}_{i} R_i$.
Evaluation of $\mathcal{U}_{\nu_1} R_i(I^{\mathrm{upd}}_{\nu_1},:)$ is
discussed in more detail in
Section~\ref{sec:HSSfact_lowrank_Schur}. When generating rows in
$R_i$, the random generator is seeded for each row using the global
row index $I_i(r)$, to ensure that $R_i$ is consistent with its
sibling. This frequent seeding is the reason the linear congruential
pseudo-random engine \verb+minstd_rand+ was chosen as default over for
instance the \verb+mt19937+ Mersenne-Twister, which has a much bigger
internal state.

The frontal matrices $\mathcal{F}_i$ with $\mathrm{level}(i) < \ell_s$
are completely approximated by HSS and are never explicitly formed as
a dense matrix. This in contrast to earlier, so-called partially
structured approaches where for instance only the $F_{11}$ or the
$F_{21}$, $F_{11}$ and $F_{12}$ blocks are
compressed~\cite{wang2014parallel}. Partially structured approaches
typically at one point or another form a dense representation of the
$F_{22}$ block, perform the Schur complement update on it, and then
use this dense update matrix in the extend-add procedure. This is done
to avoid having to perform an overly complicated extend-add operation
on HSS matrices. However, the approach followed here does not require
first assembling a dense frontal matrix before doing HSS
compression. This is due to the use of the randomized HSS compression,
Algorithm~\ref{algo:compress}, which only requires matrix-vector
multiplication and extraction of selected elements from the frontal
matrix.

When $R_i$, $S^r_i$ and $S^c_i$ have been constructed, HSS compression
using Algorithm~\ref{algo:compress} can be performed. However, when
$d_i+p$ is less than the HSS-rank of $\mathcal{F}_i$,
Algorithm~\ref{algo:compress} will fail. In that case, columns are
added to $R_i$, i.e., $d_i \mathrel{+}= \Delta d$ ($\Delta d = 128$ by
default), the new columns of $S_i^r$ and $S_i^c$ are computed and
Algorithm~\ref{algo:compress} is called again, this time only updating
the new columns of $R_i$, $S_i^r$ and $S_i^c$. Due to the use of the
ID in Algorithm~\ref{algo:compress}, HSS generators $D_\tau$ and
$B_{\nu_1,\nu_2}$ are submatrices of $\mathcal{F}_i$. Hence, a routine
to extract specific elements from $\mathcal{F}_i$ is required. This
routine will be described in Section~\ref{sec:HSS_extraction}.

\subsection{ULV factorization and low-rank Schur complement update}\label{sec:HSSfact_lowrank_Schur}
After HSS compression, a factorization of $F_{i_{11}}$ is performed:
classical row-pivoted LU if $\mathcal{F}_i$ is dense, ULV if it is
HSS.
%, see Algorithm~\ref{algo:ULVdecomposition}.  
For a dense frontal matrix, $\mathcal{U}_i = F_{i_{22}} - F_{i_{21}}
F_{i_{11}}^{-1} F_{i_{12}}$ is computed explicitly.  In the HSS case,
$F_{i_{22}}$ is kept in HSS form and the update $F_{i_{21}}
F_{i_{11}}^{-1} F_{i_{12}} = \Theta^*_i \Phi_i$ is stored as a
low-rank product. Expressions for $\Theta^*_i$ and $\Phi_i$ are
derived and presented in detail in~\cite{xia2013randomized} for
symmetric and in~\cite{xia2012superfast} for non-symmetric matrices.
% In the HSS case, $F_{i_{22}}$ is kept in HSS form and the update
% $F_{i_{21}} F_{i_{11}}^{-1} F_{i_{12}}$ is stored as a low-rank
% product.  Let $q$ be the root node of the HSS subtree corresponding
% to the $F_{i_{22}}$ block and $k$ the root node of the HSS subtree
% corresponding to $F_{i_{11}}$. Then
% \begin{equation}\label{eq:lowrankSchurupdate}
%   F_{i_{21}} F_{i_{11}}^{-1} F_{i_{12}} \equiv \Theta^*_i \Phi_i \quad \textrm{with} \quad
%   \Theta^*_i = U^{\mathrm{big}}_q B_{q,k} \hat{V}^*_k U_k^{-1} \, , \,\,
%   \Phi_i = L^{-1}_k P_k U_k B_{k,q} \left( V^{\mathrm{big}}_q \right)^* \, ,
% \end{equation}
% where $U_k$, $L_k$ and $P_k$ come from the LU factorization
% $P_kL_kU_k=\tilde{D}_k$ at HSS node $k$, see
% Algorithm~\ref{algo:ULVdecomposition},
% line~\ref{algoline:ULVrootLU}.  Derivation
% of~\eqref{eq:lowrankSchurupdate} is presented in detail in
% ~\cite{xia2013randomized} for symmetric and
% in~\cite{xia2012superfast} for non-symmetric matrices.
Given $\mathcal{U}_i = F_{i_{22}} - \Theta^*_i \Phi_i$, the
multiplication with $\mathcal{U}_i$ in~\eqref{eq:FtimesR} can be
performed efficiently using HSS matrix-vector multiplication
% Algorithm~\ref{algo:HSSmatvec} 
for $F_{i_{22}}$ and two dense (rectangular) matrix products for
$\Theta^*_i$ and $\Phi_i$.

\subsection{Extracting elements from an HSS matrix}\label{sec:HSS_extraction}
Finally, extracting elements from $\mathcal{F}_i$ requires extracting
elements from an HSS matrix. In~\cite{xia2013randomized} a routine is
presented for extracting multiple elements from an HSS matrix while
trying to minimize the number of traversals through the HSS tree. We
use a conceptually simpler algorithm based on the HSS matrix-vector
multiplication.
% , Algorithm~\ref{algo:HSSmatvec}. 
By multiplying an HSS matrix with unit vectors, selected columns can
be extracted. At the leaf nodes, instead of multiplying with a unit
vector, one can simply select the proper columns of $V^*$. Unlike for
matrix-vector multiplication, during element extraction parts of the
tree traversal can be pruned.

\ignore{
\begin{remark}
  There is a slightly faster way~\cite{xia2013randomized} to compute
  $\mathcal{U}_{\nu_1} R_{i}(I^{\mathrm{upd}}_{\nu_1},:)$, based on
  (let $k=\nu_1$):
  \begin{gather}\label{eq:indirectUR}
    \begin{bmatrix} S^r_{k;1} \\ S^r_{k;2} \end{bmatrix} = 
    \begin{bmatrix} F_{k;{11}} & F_{k;{12}} \\ F_{k;{21}} & F_{k;{22}} \end{bmatrix} 
    \begin{bmatrix} R_{k;1} \\ R_{k;2} \end{bmatrix}, \\
    \label{eq:indirectUR2}
    \mathcal{U}_k R_{k;2} = 
    \left( F_{k;{22}} - \Theta_k^* \Phi_k \right) R_{k;2} = 
    S^r_{k;2} - F_{k;{21}} R_{k;1} - \Theta_k^* \Phi_k R_{k;2} \, .
  \end{gather}
  However, this uses $S^r_k = \mathcal{F}_k R_k$, which was computed
  with $\mathcal{F}_k$ before it was approximated as HSS. Note that
  for HSS compression with $\varepsilon > 0$, $\mathcal{F}_k \approx
  \mathrm{HSSCompress}(\mathcal{F}_k)$. Since numerical experiments
  show that~\eqref{eq:indirectUR2} typically leads to larger HSS-ranks
  and uses more memory, we do not use it.
\end{remark}
}

\subsection{Preconditioning versus iterative refinement}
Direct solvers often use a few steps of iterative refinement to
improve the solution quality~\cite{wilkinson1994rounding}. However,
the multifrontal method with HSS compression as presented in this
paper is used as a preconditioner for GMRES instead. For the same
number of multifrontal solve steps (preconditioner applications), a
Krylov solver typically leads to smaller residuals than iterative
refinement. This is particularly useful when the HSS compression
tolerance is increased, since in that case the HSS-multifrontal method
is no longer an exact direct solver and the number of outer iterations
increases.
\ignore{
\todo[inline]{[FH] It's the end of the MF-HSS section but we don't cite complexity
results anywhere besides a brief sentence in the intro. Can we reuse results from Jianlin?
I would have a short 4.7 here as well as a separate 3.4 for MF instead of the current
solution+complexity section 3.3.}
}

\subsection{Solver complexity}
The computational complexity of a standard multifrontal solver is
typically dominated by the dense linear algebra corresponding to the
few largest frontal matrices. For instance a nested dissection
reordering on a d-dimensional mesh with $N = k^d$ vertices has a top
separator with $\mathcal{O}(k^{d-1})$ vertices, leading to an overall
complexity of $\mathcal{O}(k^{3(d-1)})$, i.e., $\mathcal{O}(N^{3/2})$
and $\mathcal{O}(N^2)$ for 2D and 3D meshes respectively.

For the HSS-embedded multifrontal solver, the complexity is dominated
by the HSS compression of the dense frontal matrices, which in turn
depends on the rank pattern. Earlier works by Chandrasekaran et
al.~\cite{chandrasekaran2010numerical} and Engquist \&
Ying~\cite{eng11} showed the rank patterns of the elliptic and the
Helmholtz operators respectively. Xia showed complexities for the
randomized HSS multifrontal solver assuming different rank
patterns~\cite{xia2013randomized}. Combining the above results, we
summarize the solver complexities for two types of PDEs and two sparse
solvers, see Table~\ref{tab:solver_complexity}.

\begin{table}[htp!]\footnotesize
  \centering
  % {\resizebox{\textwidth}{!}{
  \renewcommand{\arraystretch}{1.15}
  \setlength{\tabcolsep}{.2em}
  \begin{tabular}
    [c]{|c|c|c|c|c|c|c|}
    \cline{4-7}
    \multicolumn{2}{}{} & & \multicolumn{2}{c|}{MF} & \multicolumn{2}{|c|}{MF-HSS-RS}\\\hline
    \multicolumn{2}{|c|}{problem} & HSS rank
    %\multicolumn{2}{|c|}{}  &  
    & factor flops & memory & factor flops & memory\\\hline \hline
    {2D} & elliptic & $O(1)$ & \multirow{2}{*}{$O(N^{3/2})$} & \multirow{2}{*}{$O(N\log N)$}
    & \multirow{2}{*}{$O(N)$} & \multirow{2}{*}{$O(N)$}\\\cline{2-3}
    ($k\times k$)& Helmholtz & $O(\log k)$ &  &  &  & \\\hline
    {3D} & elliptic & $O(k)$ & \multirow{2}{*}{$O(N^{2})$} & \multirow{2}{*}{$O(N^{4/3})$}
    & \multirow{2}{*}{$O(N^{10/9}\log N)$} 
    & \multirow{2}{*}{$O(N)$}\\\cline{2-3}
    ($k\times k\times k$)& Helmholtz & $O(k)$ &  &  &  & \\\hline
  \end{tabular}
  % }}
  \caption{\footnotesize Summary of the complexities of the standard multifrontal solver (MF)
    and the randomized HSS-embedded multifrontal solver (MF-HSS-RS) applied to two important classes of problems.
    The mesh size per side is $k$ and the matrix dimensions are $N=k^2$
    in 2D and $N=k^3$ in 3D.}
  \label{tab:solver_complexity}
\end{table}

\section{Shared memory parallel implementation}\label{sec:implementation}
The algorithm presented in Section~\ref{sec:MFHSS} has been
implemented using C\texttt{++} and OpenMP, targeting shared memory
platforms. The code relies on BLAS, LAPACK, METIS and/or SCOTCH and a
recent C\texttt{++}11 compliant compiler with support for OpenMP 3.1
or higher. The code makes heavy use of the OpenMP task
construct. OpenMP was chosen because it is easy to use, performs well
and is well documented and supported. However, alternatives like
Intel\tm{} Threading Building Blocks~\cite{reinders2007intel} or
Cilk(\texttt{+})~\cite{blumofe1996cilk} offer conceptually similar
task parallelism. Switching to one of those should not be hard. While
other runtime systems like QUARK~\cite{yarkhan2011quark},
DAGuE/PaRSEC~\cite{bosilca2012dague} and
StarPU~\cite{augonnet2011starpu} (distributed memory task scheduling)
and OmpSs~\cite{duran2011ompss} might have certain specific advantages
over the OpenMP runtime, many of those innovations, for instance
explicit modeling of task dependencies or task-offloading, are
eventually incorporated in the OpenMP standard as well.
% (apart from the distributed memory scheduling).

OpenMP tasks are created and scheduled at runtime by the
scheduler. Task schedulers typically use a work
stealing~\cite{blumofe1999scheduling} or task stealing strategy to
balance load between threads. Each thread/core has its own local queue
of tasks. When a thread runs out of work it can steal a task from one
of the other thread's task queues.

\subsection{Task based tree parallelism}
Traversals of both the elimination tree and the HSS hierarchy allow
for tree parallelism, i.e., independent subtrees can be processed
concurrently. For instance, multifrontal factorization requires
bottom-up topological traversal of the elimination tree, just like HSS
compression requires bottom-up traversal of the HSS hierarchy. The
code in Listing~\ref{omptree} shows how to do a parallel bottom-up
tree traversal using the OpenMP task construct. The tree is stored as
objects of a class Tree with two members \verb+left_child+ and
\verb+right_child+, both pointers to subtrees, also objects of type
Tree. In Listing~\ref{omptree}, the variable \verb+depth+ keeps track
of the recursion depth and no more tasks are generated after a certain
depth to avoid excessive overhead of creating too fine-grained
tasks. Experiments show that setting \verb+d_max+ to
$\log_2(\#\textrm{threads})+3$ leads to a good task granularity. With
this setting, the maximum number of tasks at any given point in time
is about $2^{\textrm{d\_max}}=8 \cdot \#\textrm{threads}$. This is
enough to ensure good load balance and avoids excessive task creation
overhead. OpenMP tasks supports an \verb+if+ clause, so the check
\verb+if(depth<d_max)+ could have been put in the OpenMP
pragma. However, optimizing the code to perform this check outside the
directive completely avoids all task creation and synchronization
overhead when it evaluates to false. The \verb+final(condition)+
clause informs the OpenMP runtime that the generated task will not
generate more tasks if \verb+condition+ evaluates to true. Finally the
\verb+untied+ clause informs the runtime that this task can be moved
to a different thread when it encounters a scheduling point. For
instance, when a task spawns a new task, the spawning task may be
moved to another thread. Untied tasks allow for better load balance,
whereas tied tasks (the default) typically lead to better data
locality. The \verb+taskwait+ pragma ensures processing of the
children is finished before continuing with the parent.
\begin{lstlisting}[label=omptree,frame=single,captionpos=b,float,
  caption={Bottom-up topological parallel tree traversal implemented with recursion and the OpenMP~(3.1) task construct.}]
  void Tree::postorder(depth=0) {
    if (depth < d_max) {
      if (left_child)
#pragma omp task untied default(shared) final(depth >= d_max-1) mergeable
        left_child->postorder(depth+1)
      if (right_child)
#pragma omp task untied default(shared) final(depth >= d_max-1) mergeable
        right_child->postorder(depth+1)
#pragma omp taskwait
    } else {
      if (left_child) left_child->postorder(depth+1)
      if (right_child) right_child->postorder(depth+1)
    }
    do_stuff(depth); // factor/compress..., can generate more tasks
  }  
\end{lstlisting}

%\subsection{Node parallelism: parallel BLAS and LAPACK}
\subsection{Hybrid node and tree parallelism}
\begin{figure}
  \begin{subfigure}[b]{.49\textwidth}
    \includegraphics[width=\textwidth]{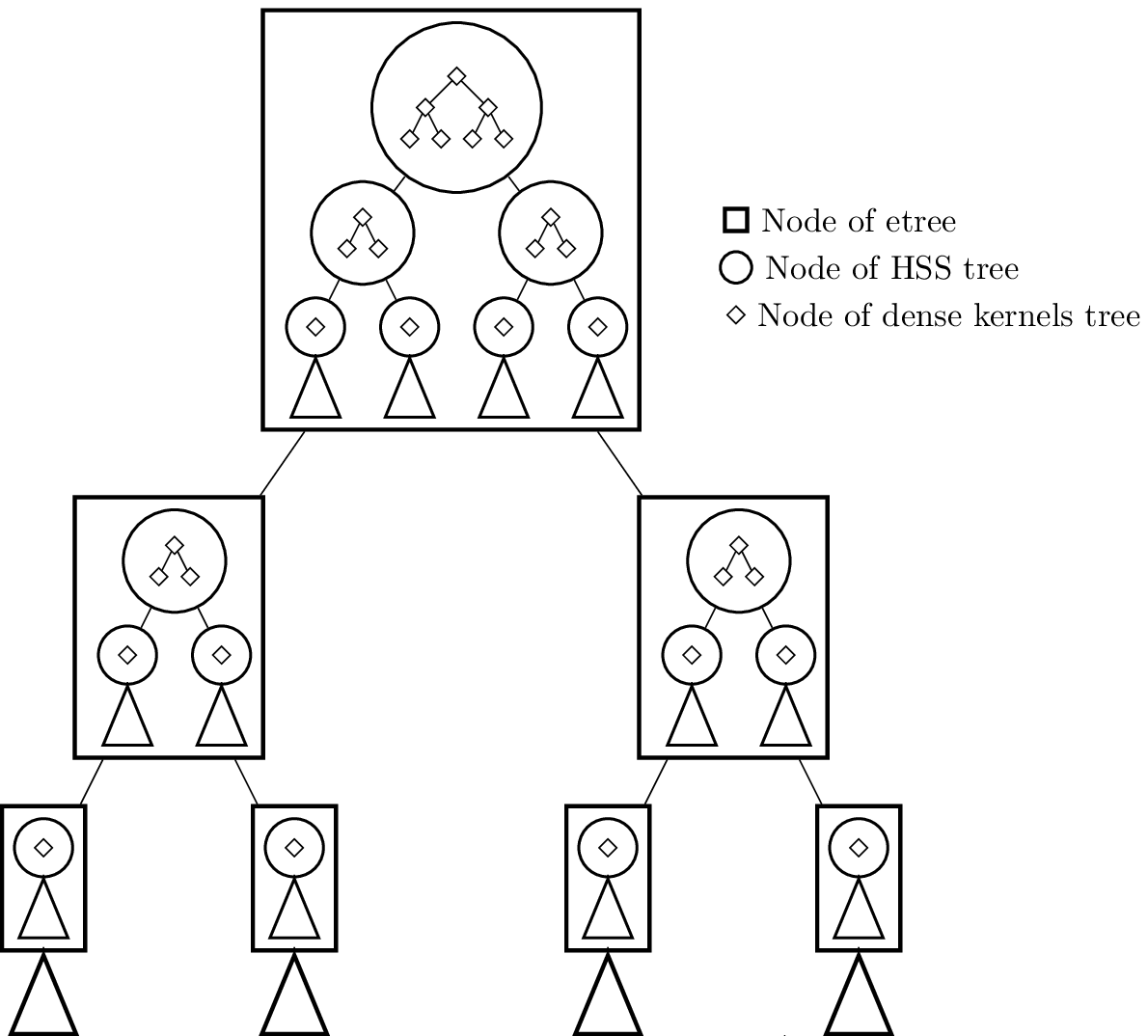}
    \vspace{.2cm}
    \caption{Three nested levels of parallelism.}
    \label{fig:dmax_tree}
  \end{subfigure}
  \begin{subfigure}[b]{.49\textwidth}
    \centering
    \includegraphics[width=\textwidth]{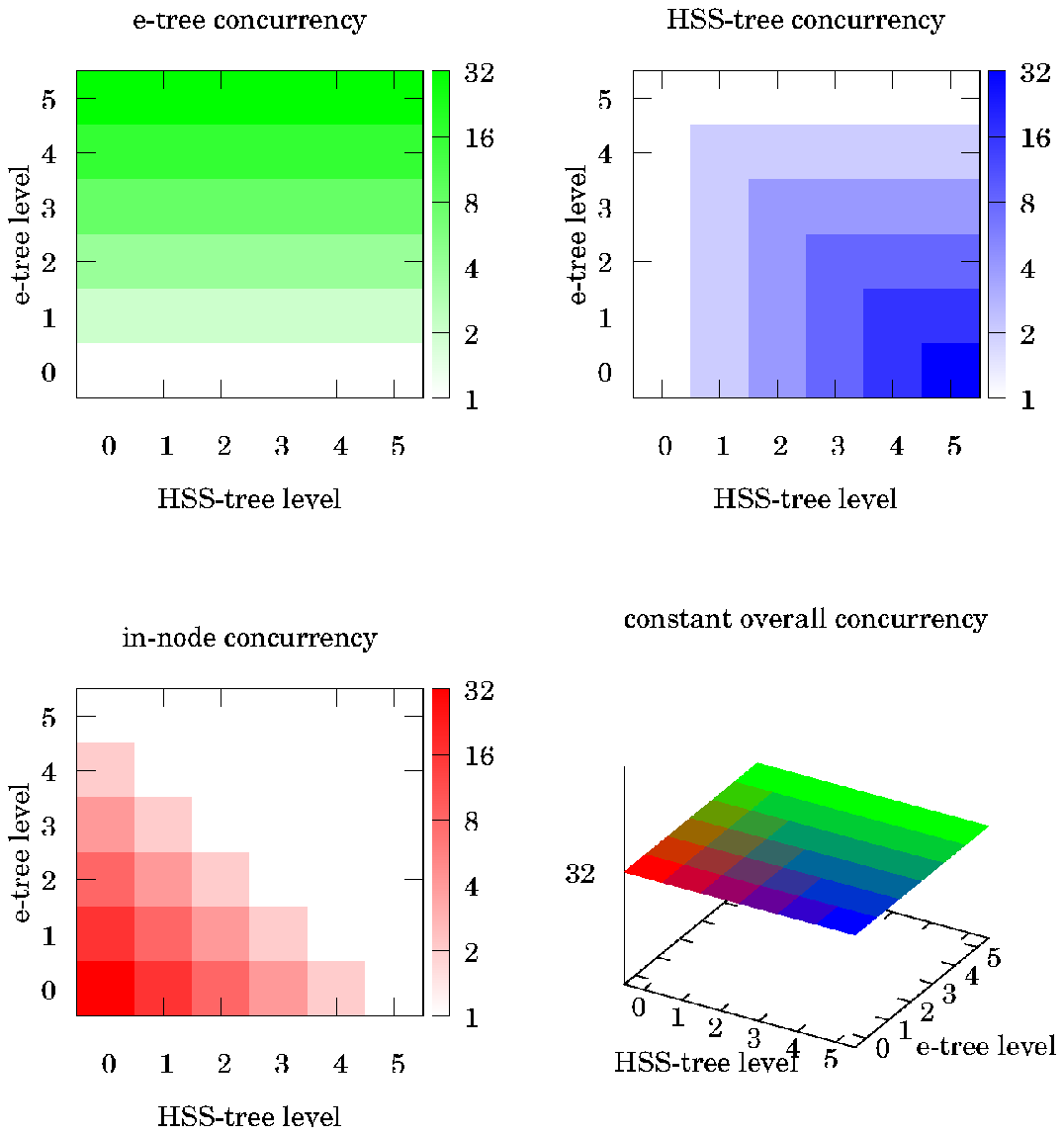}
    \caption{Constant overall concurrency.}
    \label{fig:concurrency}
  \end{subfigure}
  \caption{\footnotesize Schematic illustration of the different types
    of concurrency in the code and the gradual shift from tree
    parallelism to in-node parallelism. (a) Tasks for dense kernels
    ($\diamond$) are nested in nodes ($\circ$) of the HSS trees, which
    are nested in the elimination tree nodes ($\Box$) (e-tree). (b)
    Left-to-right, top-to-bottom: (1) Elimination tree concurrency
    decreases when getting closer to the root node. (2) Closer to the
    root of the elimination tree, more HSS tree concurrency is
    exploited as it becomes available, i.e., while moving down the HSS
    tree away from the root. (3) Towards the root of the HSS tree and
    the root of the elimination tree, more in-node concurrency
    (parallel tasked dense algebra) is exploited. (4) The product of
    the 3 types of concurrency, i.e., the overall concurrency, remains
    constant throughout both the elimination and HSS trees.}
  \label{fig:conc}
\end{figure}

Exploiting tree parallelism alone as in Listing~\ref{omptree} does not
scale well due to the limited degree of parallelism near the
root. Although the HSS-multifrontal algorithm can exploit two nested
levels of tree parallelism (elimination tree and HSS hierarchy), the
scaling bottleneck remains. To overcome this, one needs to exploit
parallelism in the computational work inside the tree nodes, which are
mostly dense linear algebra operations. However, work sharing
constructs like OpenMP \verb+parallel for+ loops are not allowed
within OpenMP tasks. Moreover, calling multithreaded BLAS or LAPACK
routines from multiple tasks/threads leads to over-subscription and
generally poor performance. This is because existing multithreaded
BLAS/LAPACK libraries are optimized to use the entire machine. One
possible strategy is to exploit tree parallelism only for the lower
levels of the tree and switch to a sequential processing of the nodes
higher up in the tree while switching to multithreaded linear
algebra. However, this leads to many synchronization points and does
not scale with increasing number of threads. Our approach on the other
hand is to use task parallelism within the tree nodes as well to allow
for a seamless transition between tree and node parallelism, since
scheduling of tasks is left to the runtime system. When getting closer
to the root node there is a shift from tree to node parallelism. This
is illustrated in Figure~\ref{fig:conc}. Even in the case of highly
unbalanced trees, the runtime can assign work evenly to the available
cores.  We chose not to use an existing library for the task based
dense linear algebra, for instance PLASMA (based on the QUARK
runtime), since we wished to exploit the same threading mechanism
(OpenMP) already used for the tree parallelism.

\subsection{Parallel BLAS and LAPACK}
% \begin{figure}
%   \begin{center}
%     \includegraphics[width=\textwidth]{figures/fig_51}
%   \end{center}
%   \caption{Speedups over the corresponding sequential Intel MKL BLAS
%     algorithm for matrix multiplication and triangular solve for
%     square matrices of size $500^2$ to $4000^2$. Left: Recursive
%     matrix multiplication implemented with OpenMP tasks. Middle:
%     multithreaded matrix multiplication from Intel MKL. Right:
%     Recursive triangular solve implemented with OpenMP tasks. }
%   \label{fig:scaling_blas}
% \end{figure}
One of the most time consuming operations of the algorithm is dense
matrix-matrix multiplication $C \gets \alpha AB+\beta C$. This can be
implemented easily with recursion and task
parallelism~\cite{mccool2012structured}, by splitting the problem into
smaller matrix-matrix multiplications; this strategy is referred to as
divide-and-conquer and is often used in so-called cache-oblivious
algorithms~\cite{frigo1999cache}. How the matrices are split depends
on their shapes. Let $A$ be $m \times k$ and $B$ be $k \times n$, then
\begin{align}\label{eq:gemmtask}
  C \gets \begin{cases}
    \alpha A B + \beta C & \textrm{if } m \times n \times k \leq T , \\
    \alpha \begin{bmatrix} A B_0 &  A B_1 \end{bmatrix} + \beta\begin{bmatrix} C_0 & C_1\end{bmatrix} & \textrm{else if } n \geq \max(m,k), \\
    \alpha \begin{bmatrix} A_0 B \\  A_1 B \end{bmatrix} + \beta\begin{bmatrix} C_0 \\ C_1\end{bmatrix} & \textrm{else if } m \geq k, \\
    \alpha \left( A_0 B_0 + A_1 B_1 \right) + \beta C & \textrm{else} \\
  \end{cases} .
\end{align}
The last case in~\eqref{eq:gemmtask}, short fat $A$ times tall skinny
$B$, uses two consecutive recursive matrix-matrix multiplication
calls. Cases 2 and 3 start two multiplications in parallel, spawning
two tasks. The recursion ends when reaching case 1, with $T$ a tuning
parameter set by default to $T=64^3$, where a sequential vendor
optimized BLAS3 \verb+*gemm+ routine is called. Depending on the
scalar type, one of four inlined template specialization functions for
\verb+gemm<scalar>+ is executed to pick the correct version:
\verb+sgemm+, \verb+dgemm+, \verb+cgemm+ or \verb+zgemm+. For the
other BLAS2/3 routines that are required, for instance triangular
matrix multiplication and solve, a similar recursive approach is
used. This recursive task generation is also stopped when the
recursion depth becomes too large, with the same \verb+depth+
parameter being passed through the entire code and incremented each
time it enters a new task.

The code also requires some LAPACK functionality, namely LQ, LU and
RRQR decompositions. For those, we modify the reference Fortran LAPACK
implementation to make use of our parallel (tasked) BLAS
routines. Some vendor optimized LAPACK libraries do not just use the
LAPACK reference code on top of multithreaded BLAS calls, but add
additional optimizations to the LAPACK routines. Unfortunately, in our
approach we cannot take advantage of these optimized multithreaded
codes.
\begin{figure}
  \begin{center}
    \includegraphics[width=\textwidth]{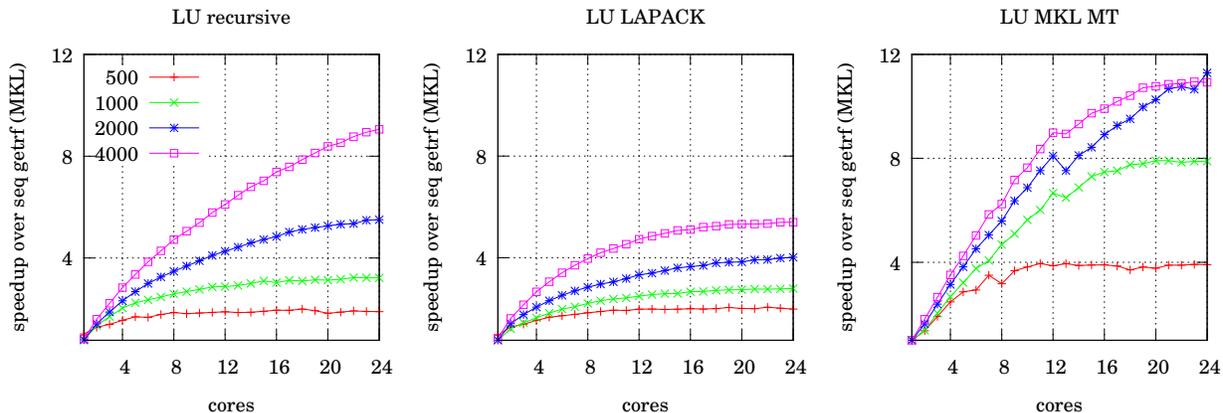}
  \end{center}
  \caption{\footnotesize Speedups over sequential \texttt{getrf} from
    Intel\tm{} MKL for matrices of size $500^2$ to $4000^2$. Left:
    Recursive LU decomposition using OpenMP tasked BLAS code. Middle:
    Reference LAPACK \texttt{getrf} using OpenMP tasked BLAS
    code. Right: MKL optimized multithreaded \texttt{getrf}. Our
    recursive implementation scales better than the reference netlib
    \texttt{getrf} with parallel BLAS but worse than the MKL optimized
    code. However, calling MKL multithreaded \texttt{getrf} from
    multiple threads simultaneously would lead to over-subscription
    and performance penalty. This is not a problem with the recursive
    LU because it uses the OpenMP task runtime, just like the rest of
    the code.}
  \label{fig:scaling_LU}
\end{figure}
Consider partial pivoted LU decomposition, used for the $F_{11}$ block
of a dense frontal matrix. Apart from the LAPACK \verb+*getrf+ routine
using our OpenMP tasked BLAS routines, we also implemented a recursive
LU factorization
algorithm~\cite{recursiveTileLU}. % , which works as follows. Partition
% the matrix in two column blocks, recursively compute a partial
% pivoting LU for the left block, apply the permutation and the $U$
% factor to the right block, do a Schur complement update of the lower
% right block, compute an LU decomposition of the lower right block and
% finally apply the permutation of that last LU decomposition to the
% lower left block.
The parallelism in this algorithm has to come from the BLAS routines,
triangular solve, row permutation, and matrix-matrix
multiply. Figure~\ref{fig:scaling_LU} compares the performance and
scalability of the two LU decomposition approaches with the MKL
optimized implementation and shows our implementation of LU scales
nearly as well as MKL without sacrificing the ability to exploit
subtree concurrency. A more scalable approach~\cite{buttari2009class},
based on so-called tiled algorithms instead of recursion, partitions
the matrices in tiles of fixed sizes and assigns tasks to each of the
tiles while explicitly modeling the data dependencies between the
tasks. A DAG (directed acyclic graph) scheduler then executes the
tasks while respecting their dependencies. OpenMP supports explicit
task dependencies since version 4.0\footnote{Not all compilers
  currently support the latest OpenMP 4.0 standard.}. We intend to
exploit this feature in the future to achieve more scalable dense
linear algebra operations.
% \begin{figure}
%   \begin{center}
%     \includegraphics[width=\textwidth]{figures/fig_53}
%   \end{center}
%   \caption{TODO Figure and caption need to be redone! Scaling of two
%     RRQR decomposition implementations.}
%   \label{fig:scaling_RRQR}
% \end{figure}
% Also for the rank-revealing QR decomposition we have two
% implementations. One is a modified version of the LAPACK
% \verb+*geqp3+ code~\cite{quintana1998blas}, a BLAS3 version of
% column pivoted QR.
For the rank-revealing QR decomposition we use a modified version of
the LAPACK \verb+*geqp3+ code~\cite{quintana1998blas}, a BLAS3 version
of column pivoted QR. The routine is modified to call our parallel
tasked BLAS and an extra tolerance parameter $\varepsilon$ is added to
stop the rank-revealing process as soon as the $\varepsilon$-rank has
been found instead of computing the full decomposition. More
precisely, numerical rank $i$ is detected when $R_{i+1,i+1} / R_{11}
\leq \varepsilon$, where $R$ is the upper-triangular factor.

\subsection{Scaling bottlenecks}
Before the actual numerical factorization step, but after matrix
scaling and nested dissection reordering, a symbolic factorization
step is performed. During this step some memory is allocated and the
index sets $I^{\mathrm{upd}}_{\tau}$ are assembled.
% Index set $I^{\mathrm{upd}}_{\tau}$ for node $\tau$ is the
% combination of the index sets of its children.
The symbolic factorization is a bottom-up tree traversal which is done
in parallel, as in Listing~\ref{omptree}.
% For a pure multifrontal method all memory for the factors can be
% allocated during the symbolic factorization. In a sequential
% multifrontal code, one can also compute the maximum required amount of
% temporary storage for the update blocks during the numerical
% factorization and allocate once at the beginning of the factorization.
% For a sequential purely multifrontal method without delayed pivoting,
% the peak memory usage can be determined during the symbolic
% factorization step and memory can be allocated in a single big chunk.
% However, in a dynamically scheduled parallel setting, this is much
% harder, so our code simply requests and frees memory when
% required. Similarly, the memory for the HSS generators of the
% multifrontal factors is only allocated when the HSS-rank has been
% determined. 
In a multithreaded setting, memory allocation can become a serious
scaling bottleneck. We have found that the use of a scalable memory
allocator, like TCMalloc~\cite{ghemawat2009tcmalloc} or the TBB
scalable memory allocator~\cite{reinders2007intel} greatly improves
the performance over for instance the default \verb+malloc+ in
glibc\footnote{\url{http://www.gnu.org/software/libc/}}. For instance,
running on a 60 core Intel\tm{} Xeon Phi, the symbolic factorization
phase runs up to $56 \times$ faster when using TBBMalloc instead of
the default allocator.

% Allocating memory when it is needed instead of initially allocating
% one large chunk of memory helps keep the code simple and it improves
% data locality especially on NUMA architectures.

\section{Numerical experiments}\label{sec:experiments}
This section presents various numerical
results. Section~\ref{sec:exp_pde} first focuses on some PDE problems
on regular grids as this allows us to easily change the problem
size. The following sections consider other matrices from various
applications.  Unless otherwise stated, the experiments are performed
on a single $12$-core socket of a single node of the NERSC Edison
machine\footnote{\url{https://www.nersc.gov/users/computational-systems/edison/}}. A
compute node has two $12$-core Intel\tm{} Ivy Bridge processors at
$2.4$GHz. Double precision peak performance is $19.2$Gflop/s per core,
$230.4$Gflop/s per socket or $460.8$Gflop/s per node. Each socket has
$32$GB DDR3 $1866$MHz memory, hence $64$GB per node, with a
STREAM~\cite{streambench} bandwidth of $48.5$GB/s. We use the
Intel\tm{} 15.0.1 compiler with sequential MKL.

% When using $12$ or less threads, all threads are bound to a single
% socket and only $32$GB is available.

\subsection{PDEs on a regular grid}\label{sec:exp_pde}
We start with a number of benchmarks of well-known PDEs on regular 2D
and 3D grids to study scaling of time-to-solution, number of floating
point operations, memory usage, HSS-ranks etc., with respect to
problem size. For these regular grids, a geometric nested dissection
code is used instead of the default METIS graph partitioner. The
following benchmark problems are considered:
\begin{itemize}
\item Poisson equation $-\Delta u = f$ on a 2D grid using the standard
  $5$-point finite difference stencil with homogeneous Dirichlet
  boundary conditions.
\item Poisson equation on a 3D grid using the standard $7$-point
  stencil with homogeneous Dirichlet boundary conditions.
\item Convection diffusion equation~\cite{agmg} $-\nu \Delta u +
  \mathbf{v} \cdot \nabla u = f$ on a 2D grid using a $5$-point upwind
  stencil, with viscosity $\nu = 10^{-4}$ and
  \begin{equation}
    \mathbf{v} = \begin{pmatrix} x(1-x)(2y-1) & y(1-y)(2x-1)\end{pmatrix}^T \, .
  \end{equation}
\item Convection diffusion, similar as above, on 3D grid with
  \begin{equation}
    \mathbf{v} = \begin{pmatrix}  2x(1-x)(2y-1)z & -y(1-y)(2x-1) & -(2x-1)(2y-1)z(1-z) \end{pmatrix}^T \, .
  \end{equation}
\item Helmholtz equation
  \begin{equation}
    \left( -\Delta - \omega^2 / v(x)^2 \right) u(x,\omega) = s(x, \omega)
  \end{equation}
  on a 2D grid, with $\omega$ the angular frequency, $v(x)$ the
  seismic velocity and $u(x,\omega)$ the time-harmonic wavefield
  solution to the forcing term $s(x,\omega)$. The discretization uses
  a $9$-point stencil and the frequency is set at $f = 10$Hz with
  $\omega = 2 \pi f$. We use a sampling rate of about $15$ points per
  wavelength and PML boundary conditions. This example uses complex
  arithmetic.
\item Same as H2D, but 3D using a $27$-point stencil.
\end{itemize}

A crucial parameter for performance is the number of levels $\ell_s$
of the elimination tree for which HSS compression is performed. Note
that $\ell_s = 0$ corresponds to a pure multifrontal
solver. Unfortunately, the optimal $\ell_s$ is impossible to predict a
priori, so it is determined experimentally and will always be
mentioned with each result. The same applies to the compression tolerance
$\varepsilon$. % The observed optimal values of $\ell_s$ and
% $\varepsilon$ for the benchmark problems used in this section should
% help in finding good values for other problems as well.
% In~\cite{xia2013randomized}, Xia suggests to set $\ell_s$ such that
% the cost of factorization for the levels above the switch-level
% $\ell_s$ equals the cost of factorization for the levels below the
% switch-level $\ell_s$. However, since the HSS-ranks are not known a
% priori, this is not a practical guideline.  The code sets default
% values $\ell_s = 0$ and $\varepsilon = 10^{-4}$.
When $\ell_s > 0$, i.e., with HSS compression, the multifrontal solver
is used as a preconditioner for restarted GMRES($30$) with modified
Gram-Schmidt and a zero initial guess. Without HSS compression,
iterative refinement with the direct solver is used. All experiments
are performed in double precision with relative or absolute stopping
criteria $\| u_i\| / \|u_0\| \leq 10^{-6}$ or $\| u_i\| \leq
10^{-10}$, where $u_i = M^{-1}(A x_i - b)$, with $M$ the approximate
multifrontal factorization of $A$, is the preconditioned residual. The
right-hand-side is always set to $A \begin{bmatrix} 1 & 1 & \cdots &
  1 \end{bmatrix}^T$.

\begin{figure}
  \begin{center}
    \begin{subfigure}{.49\textwidth}
      \includegraphics[width=\textwidth]{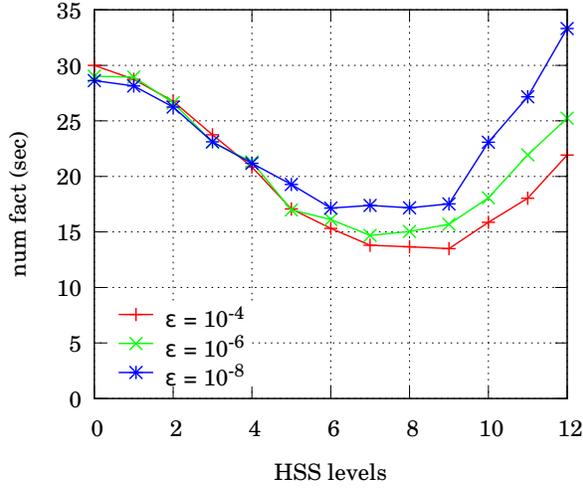}
      \caption{\footnotesize Num fact time, $5000^2$.}
      \label{fig:61}
    \end{subfigure}
    \begin{subfigure}{.49\textwidth}
      \includegraphics[width=\textwidth]{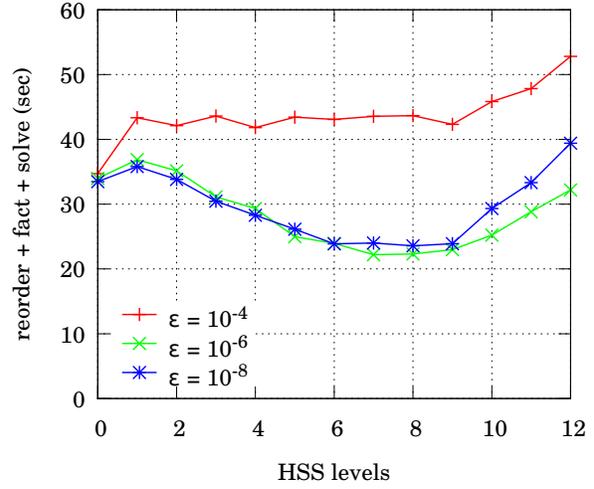}
      \caption{\footnotesize Total solve time, $5000^2$.}
      \label{fig:62}
    \end{subfigure}
    \begin{subfigure}{.49\textwidth}
      \includegraphics[width=\textwidth]{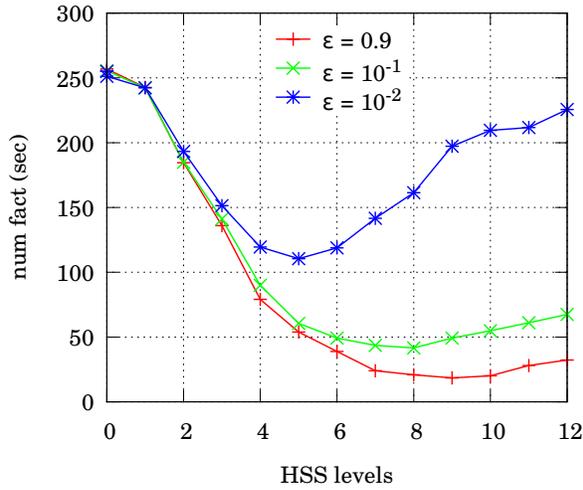}
      \caption{\footnotesize Num fact time, $125^3$.}
      \label{fig:63}
    \end{subfigure}
    \begin{subfigure}{.49\textwidth}
      \includegraphics[width=\textwidth]{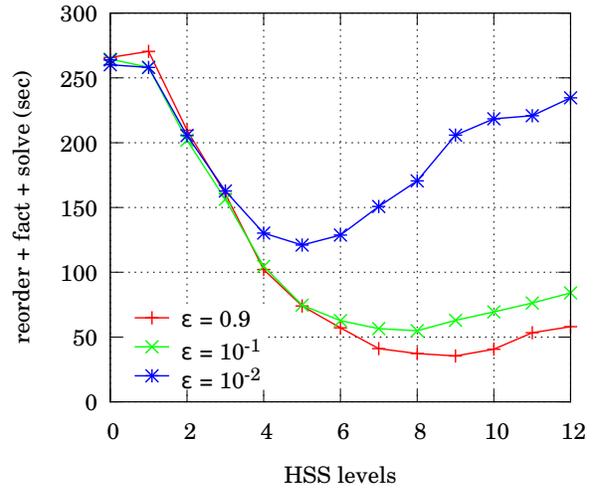}
      \caption{\footnotesize Total solve time, $125^3$.}
      \label{fig:64}
    \end{subfigure}
  \end{center}
  \caption{\footnotesize Times for factorization (left) and solve (right) of
    the 2D (top) and 3D (bottom) Poisson equation on $5000^2$ and
    $125^3$ grids as function of the number of levels $\ell_s$ in the
    elimination tree for which HSS compression is applied. Different
    curves correspond to different HSS compression tolerances
    $\varepsilon$. For 3D, much more aggressive HSS compression can be
    used.}
  \label{fig:Poisson2D3D}
\end{figure}
Figure~\ref{fig:Poisson2D3D} shows timing results for the 2D~(top) and
3D~(bottom) Poisson equation on $5000^2$ and $125^3$ grids
respectively. Figure~\ref{fig:61} shows numerical factorization time
as a function of the number of levels in the elimination tree for
which HSS compression was used. The HSS levels always correspond to
the top levels of the elimination tree. This shows that applying HSS
compression leads to a speedup of about $2 \times$ for $7$ HSS
levels. Different lines correspond to different HSS compression
tolerances $\varepsilon$. Somewhat larger factorization speedups are
possible for $\varepsilon \geq 10^{-4}$. However, this does not lead
to faster time-to-solution. Figure~\ref{fig:62} shows the cumulative
time for nested dissection reordering, symbolic factorization,
numerical factorization and GMRES solve. For $\varepsilon \geq
10^{-4}$, the number of GMRES iterations, and thus the number of
applications of the multifrontal solve, increases too much to get
overall speedup. Best results were obtained with $\ell_s = 8$,
$\varepsilon = 10^{-7}$ and only $2$ GMRES iterations ($3$
multifrontal solves). Figures~\ref{fig:63} and~\ref{fig:64} show the
timings for the 3D Poisson problem. For the 3D problem, much more
aggressive HSS compression can be used. Best results were obtained
with $\ell_s= 10$, $\varepsilon = 0.9$ and $61$ GMRES iterations. For
the Poisson problem it seems that for 2D the direct solver is very
efficient, with a modest speedup from HSS, while for 3D the HSS
enabled factorization leads to a good preconditioner.

\begin{figure}
  \begin{center}
    \begin{subfigure}{.49\textwidth}
      \includegraphics[width=\textwidth]{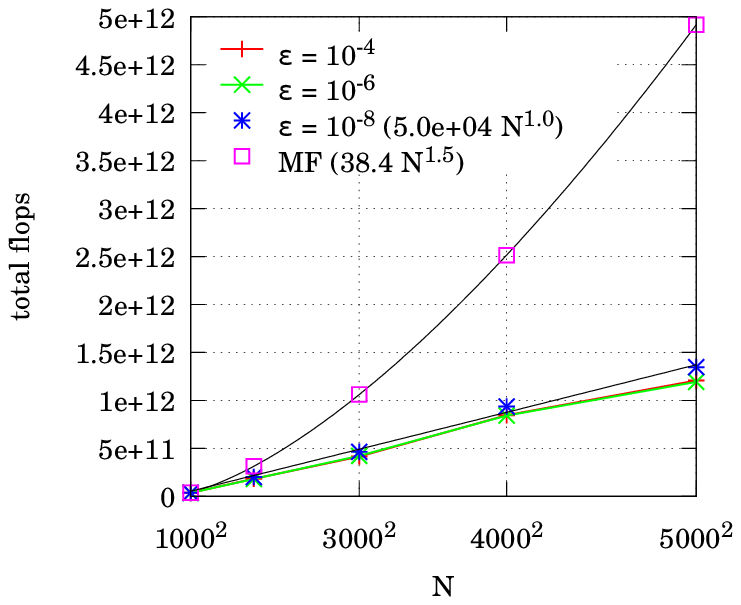}
      \caption{\footnotesize 2D Poisson.}
      \label{fig:67}
    \end{subfigure}
    \begin{subfigure}{.49\textwidth}
      \includegraphics[width=\textwidth]{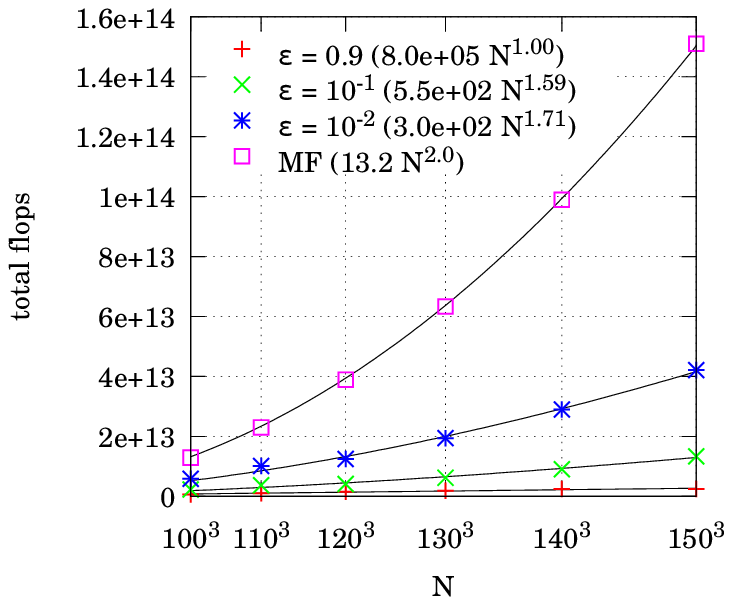}
      \caption{\footnotesize 3D Poisson.}
      \label{fig:610}
    \end{subfigure}
  \end{center}
  \caption{\footnotesize Scaling of the number of floating point operations
    required to factor and solve a 2D(a) or 3D(b) Poisson
    equation. (a) The theory predicts $\mathcal{O}(N^{3/2})$
    complexity for the multifrontal (MF) solver and optimal
    $\mathcal{O}(N)$~\cite{xia2013randomized} complexity with HSS
    compression. (b) $\mathcal{O}(N^{2})$ complexity for the
    multifrontal solver and slightly lower complexity with HSS
    compression. The fits (black lines) are very sensitive to the data
    and not very reliable. However, note the smaller exponents and the
    larger constants for the new solver.}
  \label{fig:Poisson2D3D_scaling}
\end{figure}
Figure~\ref{fig:67} shows the total number of flops (numerical
factorization and GMRES solve) for solving a 2D Poisson equation as
function of the number of degrees of freedom, again for different
compression tolerances. For the pure multifrontal method (no
compression), the number of flops is $\mathcal{O}(N^{3/2})$, as
predicted by the theory. For 2D Poisson the HSS-rank is independent of
the grid size~\cite{chandrasekaran2010numerical}, which leads to an
optimal solver, i.e., linear scaling in the number of unknowns, see
the fit in Figure~\ref{fig:67}.
% The fit in Figure~\ref{fig:67} for $\varepsilon =
% 10^{-8}$ shows an exponent even slightly smaller than one, which is
% probably due to the fact that for the smaller problems too many
% random vectors are used (oversampling).
Note the much larger constant for the HSS method. For the $5000^2$
problem there is a reduction in the number of flops by a factor of
about $3.3\times$. However, the observed speedup (Figure~\ref{fig:62}) is
smaller than that. This is due to the fact that although the number of
flops for the factorization decreases, the number of flops for the
solution phase (and GMRES iterations) increases. Although multifrontal
solve requires an order of magnitude less flops than factorization, it
runs at much lower flop rates on modern hardware because it is limited
by the memory bandwidth instead of the floating point
unit. Additionally, the flop rate in the factorization phase is lower
when using HSS compression due to the more fine-grained task
decomposition.
% since the algorithm is more complex than pure multifrontal.
Figure~\ref{fig:610} shows number of flops-to-solution for the 3D
Poisson equation. For the very aggressive compression,
$\varepsilon=0.9$, the number of floating point operations for the
$125^3$ problem are reduced to $4.4\%$ of the number of flops for the
multifrontal method.
% The data for Figure~\ref{fig:610}, up to $N=150^3$, was obtained in
% single precision.
\begin{figure}
  \begin{center}
    \begin{subfigure}{.49\textwidth}
      \includegraphics[width=\textwidth]{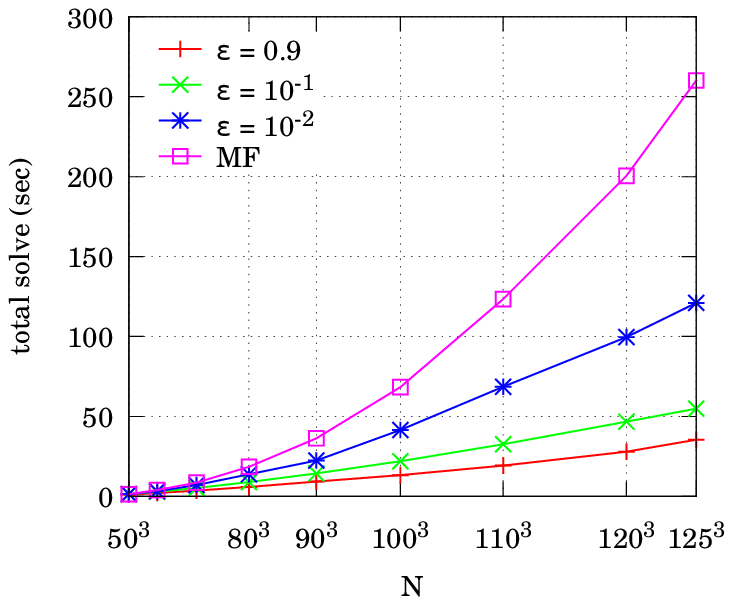}
      \caption{\footnotesize 3D Poisson, total time.}
      \label{fig:612}
    \end{subfigure}
    \begin{subfigure}{.49\textwidth}
      \includegraphics[width=\textwidth]{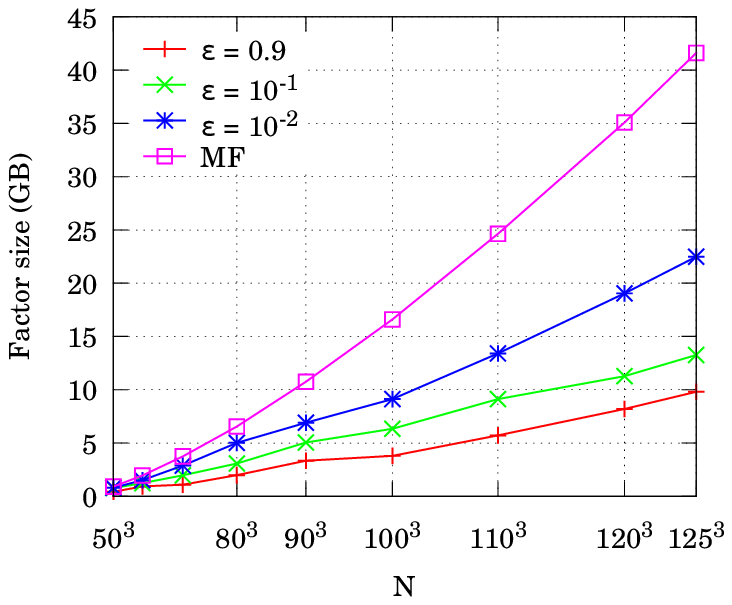}
      \caption{\footnotesize 3D Poisson, factor size.}
      \label{fig:613}
    \end{subfigure}
  \end{center}
  \caption{\footnotesize (a) Total solve time (reordering, factorization and
    solution) for the 3D $125^3$ Poisson equation. (b) Factor size for
    the same problem. Different lines correspond to different HSS
    compression tolerance $\varepsilon$, MF refers to pure
    multifrontal. The HSS enabled solver is faster for larger
    problems, and it allows to solve larger problems.}
  \label{fig:Poisson3D_time_mem}
\end{figure}
Figure~\ref{fig:612} shows the total solve time for the 3D problem for
different grid sizes. These times include matrix reordering,
factorization and GMRES solve. Figure~\ref{fig:613} shows the size of
the factors.

\begin{figure}
  \begin{center}
    \includegraphics[width=\textwidth]{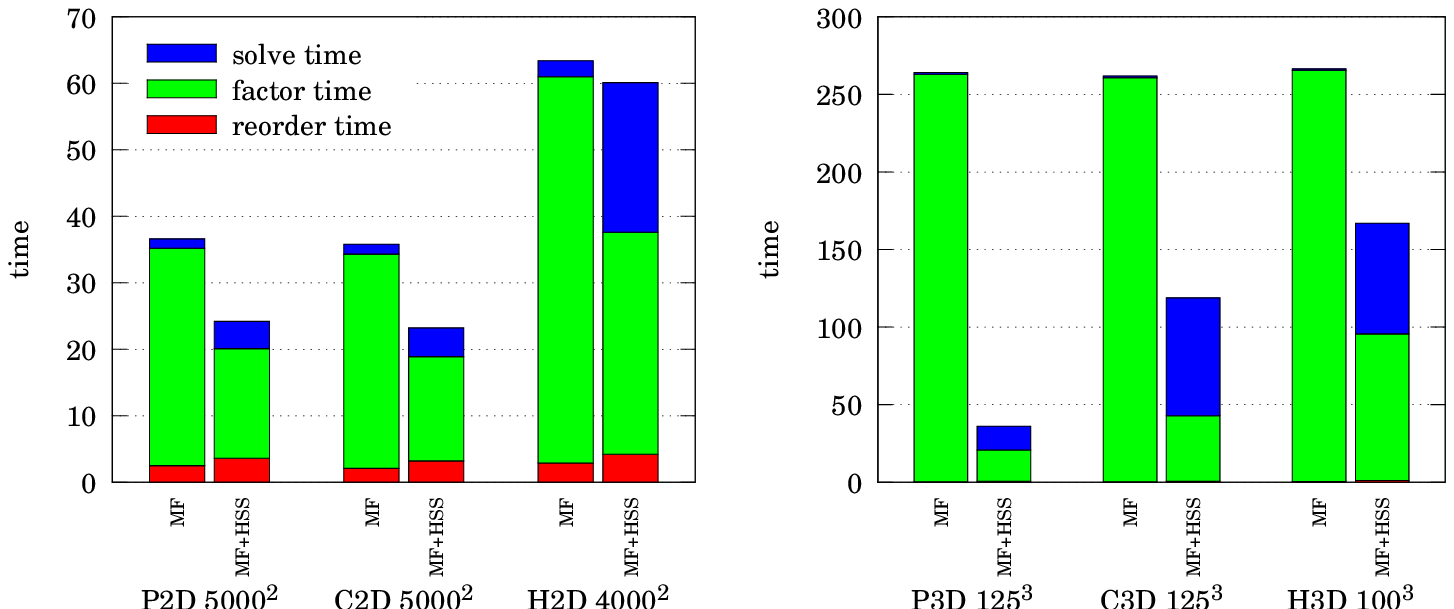}
  \end{center}
  \caption{\footnotesize Summary of the results from
    Table~\ref{tab:pde_table}: Poisson (P), convection-diffusion (C)
    and Helmholtz (H) on 2D (left) and 3D (right) regular grids on a
    $12$-core Intel\tm{} Ivy Bridge. Poisson and convection-diffusion
    are in double precision, Helmholtz in complex double precision.}
  \label{fig:hist_pde}
\end{figure}

\begin{table}[htp!]\footnotesize
  \renewcommand{\arraystretch}{1.25}
  \setlength{\tabcolsep}{0.35em}
  \newcommand{\ce}[1]{\multicolumn{1}{c|}{#1}}
  \newcommand{\cee}[1]{\multicolumn{1}{c|}{#1}}
  \begin{center}
    \begin{tabular}{|c|l||c|c|c|c|c|c|}
      \hhline{|-|-|-|-|-|-|-|-|}
      \multicolumn{2}{|r||}{problem}                                                                          
      &    \cee{P2D} &    \cee{P3D} &    \cee{C2D} &   \cee{C3D}  &    \cee{H2D} &   \cee{H3D} \\ 
      \hhline{|-|-|-|-|-|-|-|-|}
      \multicolumn{2}{|r||}{grid size}
      &        $5000^2$ &      $125^3$ &    $5000^2$ &     $125^3$ &  $4000^2$ &     $100^3$ \\ 
      \hhline{==::=:=:=:=:=:=}
      \multirow{10}{*}{\rotatebox{90}{Multifrontal}}                                                 
      &              nested dissection time (s) &  2.5    &  0.25  & 2.1   &  0.24   & 2.9  & 0.43  \\ 
      \hhline{|~-||-|-|-|-|-|-|}                                                         
      &         symbolic factorization time (s) &  3.6    &  8.0   &  3.4  &  8.1   & 4.5  &  6.0   \\ 
      \hhline{|~-||-|-|-|-|-|-|}                                                         
      &                  factorization time (s) &  29.1  &  254.7  &  28.8 &  252.4 & 53.6 &  259.1 \\ 
      \hhline{|~-||-|-|-|-|-|-|}                                                         
      &    factorization flops ($\times10^{12}$) &   4.9  &  50.0  &  4.9  &   50.0  & 10.1  &  53.5  \\ 
      &   \qquad flop rate ($\times10^9$Gflop/s) & 168.4 & 196.3   & 170.1 & 198.1  & 188.4  & 206.5 \\
      &   \qquad fraction of peak                & 73\%  & 85\%    & 74\%  & 86\%   & 82\%   & 90\% \\
      \hhline{|~-||-|-|-|-|-|-|}                                                         
      &                        factor size (GB) &  28.4   &  41.6  &  28.4  & 41.6  & 36.3   &  35.5 \\ 
      \hhline{|~-||-|-|-|-|-|-|}                                                         
      &                       solution time (s) &   1.4    & 1.1   &  1.5  &  1.1   & 2.4   &  0.91  \\ 
      \hhline{|~-||-|-|-|-|-|-|}                                                         
      &          solution flops ($\times10^{9}$) &  7.9    &  10.5  &  7.9  & 10.5  &  21.1  &  18.2  \\ 
      \hhline{|~-||-|-|-|-|-|-|}                                                         
      &                solution bandwidth (GB/s) &  20.3  &  37.8  & 18.9  & 37.8   &  15.1  & 39.0  \\ 
      &   \qquad fraction of peak                & 42\%   & 78\%   & 39\% & 78\%    & 31\%   & 80\% \\
      \hhline{|~-||-|-|-|-|-|-|}                                                               
      &          total flops ($\times10^{12}$) &  4.9     & 50.0   & 4.9    &  50.0  & 10.1 & 53.5 \\ 
      \hhline{|~-||-|-|-|-|-|-|}                                                               
      &                         total time (s) &  36.6   & 264.1  &  35.8   & 261.8  & 63.4 & 266.4 \\ 
      \hhline{==::=:=:=:=:=:=}
      \multirow{11}{*}{\rotatebox{90}{Multifrontal + HSS}}                                               
      &              nested dissection time (s) &   2.5   & 0.26    &  2.1   & 0.24  &  2.9  &  0.43 \\ 
      \hhline{|~-||-|-|-|-|-|-|}                                                               
      &            separator reordering time (s) &  1.1   &   0.40   &  1.1   & 0.35   &  1.3  &  0.68 \\ 
      \hhline{|~-||-|-|-|-|-|-|}                                                               
      &         symbolic factorization time (s) &   2.0   &  0.55    &  2.1   &  0.8  &  2.9   &  1.8 \\ 
      \hhline{|~-||-|-|-|-|-|-|}                                                               
      &                  factorization time (s) &  14.5   &  19.6   &  13.6  & 41.5  &  30.5  &  92.8 \\ 
      \hhline{|~-||-|-|-|-|-|-|}                                                               
      &    factorization flops ($\times10^{12}$) &  1.5   &  2.0    &  1.3   &  5.0   &  3.9  & 18.0  \\ 
      &   \qquad flop rate ($\times10^9$Gflop/s) &  103.4 &  102.0  &  95.6  &  120.5 & 127.9 & 194.0 \\
      &   \qquad fraction of peak                &  45\%  &  44\%   &  41\%   & 52\%   &  56\% &  84\% \\
      \hhline{|~-||-|-|-|-|-|-|}                                                               
      &                        factor size (GB) &   22.5   &   9.8  &  21.6  & 14.6   &  29.6 & 21.4 \\ 
      &      \qquad fraction of multifrontal      &  \textbf{79\%}   & \textbf{24\%}    & \textbf{76\%}   & \textbf{35\%}   & \textbf{82\%}  & \textbf{60\%} \\
      \hhline{|~-||-|-|-|-|-|-|}                                                               
      &                       solution time (s) &   4.1   & 15.3    &  4.3   &  75.9  & 22.5  &  71.2 \\ 
      \hhline{|~-||-|-|-|-|-|-|}                                                               
      &                      GMRES(30) iterations  &   3     &   67    &  3     &  234  &  11   &  152 \\ 
      \hhline{|~-||-|-|-|-|-|-|}                                                               
      &          solution flops ($\times10^{9}$) &  25.6   &  169.9   &  23.7 &  876.3 &  210.5 & 1,759.0 \\ 
      \hhline{|~-||-|-|-|-|-|-|}                                                         
      &                solution bandwidth (GB/s) & 22.0  & 43.6  & 20.1  & 45.2  & 15.8 & 46.0  \\
      &   \qquad fraction of peak                &  45\% & 90\%  &  41\%  & 93\%  & 33\% & 95\% \\
      \hhline{|~-||-|-|-|-|-|-|}                                                               
      &             HSS levels $\ell_s$ (total) &  7 (22)  &  8 (18) &  8 (22) & 7 (18) & 7 (22) &  4 (18) \\ 
      \hhline{|~-||-|-|-|-|-|-|}                                                               
      &                                HSS-rank &  48      &   46    &  50    &  397   &  139  & 30 \\ 
      \hhline{|~-||-|-|-|-|-|-|}                                                               
      & HSS compression tolerance $\varepsilon$ & $10^{-6}$ & 0.9 &  $10^{-5}$ &  0.1  & $10^{-4}$ & 0.9 \\ 
      \hhline{|~-||-|-|-|-|-|-|}                                                               
      &          total flops ($\times10^{12}$) &  1.5    &  2.2 & 1.3 & 5.9 & 4.1 & 19.8    \\ 
      &    \qquad fraction of multifrontal     &  \textbf{30\%} & \textbf{4.4\%} & \textbf{27\%} & \textbf{12\%} & \textbf{41\%} & \textbf{37\%}  \\
      \hhline{|~-||-|-|-|-|-|-|}                                                               
      &                         total time (s) &  24.2   &  36.1  &  23.2  &  118.8  & 60.1 & 166.9 \\ 
      \hhline{|~-||-|-|-|-|-|-|}                                                               
%      &   \qquad fraction of multifrontal     &  \textbf{66\%}   & \textbf{14\%}  & \textbf{65\%} & \textbf{45\%} & \textbf{95\%} & \textbf{63\%}    \\
      &   speedup                              &  \textbf{1.52$\times$}  & \textbf{7.14$\times$}  & \textbf{1.54$\times$} & \textbf{2.22$\times$} & \textbf{1.05$\times$} & \textbf{1.59$\times$}    \\
      \hhline{|--||-|-|-|-|-|-|}
    \end{tabular}
  \end{center}
  \caption{\footnotesize Comparison of the standard multifrontal solver and the multifrontal
    solver with HSS compression for a number of PDEs on regular
    grids. All experiments are run on a $12$-core Intel\tm{} Ivy Bridge (peak
    $230.4$Gflop/s and $48.5$GB/s) in double
    precision. The code achieves good performance in terms of Gflop/s (for the
    factorization) or GByte/s (for the solve) and HSS compression leads to
    nice speedups over the standard multifrontal solver.
    A geometric nested dissection code is used for these regular grid problems.}
  \label{tab:pde_table}
\end{table}
Table~\ref{tab:pde_table} shows detailed results for the 6 PDE
problems. The best speedups are obtained for the 3D problems. The code
achieves good performance in flops per second for the factorization
phase; although slightly less so for the HSS enabled code. Since the
performance of the solve phase is not bounded by the floating point
unit but rather by the memory bandwidth, we report the approximate
attained bandwidth.
% , computed as factor size times number of
% applications of the preconditioner divided by solve time. The number
% of preconditioner applications was always 1 for the multifrontal
% solver, i.e., no iterative refinement was necessary. With HSS
% compression, the number of preconditioner applications is the number
% of GMRES iterations plus 1. Especially for the 3D problems the
% attained bandwidth is very close the maximum as measured with the
% STREAM benchmark.
The detailed results from Table~\ref{tab:pde_table} are summarized in
Figure~\ref{fig:hist_pde}.

The $\varepsilon$ and $\ell_s$ values used for
Table~\ref{tab:pde_table} and
Figures~\ref{fig:Poisson2D3D}-\ref{fig:Poisson3D_time_mem} were chosen
to minimize the total time to factor and solve a single linear system,
i.e., the optimal trade-off between factorization time and number of
GMRES iterations. When multiple consecutive solves with the same
matrix are required, one needs to select different $\ell_s$ and
$\varepsilon$ values. For many consecutive and highly accurate solves,
the pure (exact) multifrontal factorization is probably optimal as it
minimizes the number of multifrontal triangular solves. However,
suppose only a few digits of accuracy are required. The multifrontal
HSS solver can be used as a direct solver and due to the smaller
factor size the solve phase will be faster than a solve with the pure
multifrontal code.

The times for symbolic factorization in Table~\ref{tab:pde_table} are
larger for the multifrontal method than for the HSS solver. This is
because the memory for dense frontal matrices is allocated during the
symbolic factorization while memory for the HSS generators is
allocated during the numerical factorization since HSS-ranks are not
known in advance.

\subsection{Matrices from various applications}
\begin{figure}
  \begin{center}
    \includegraphics[width=\textwidth]{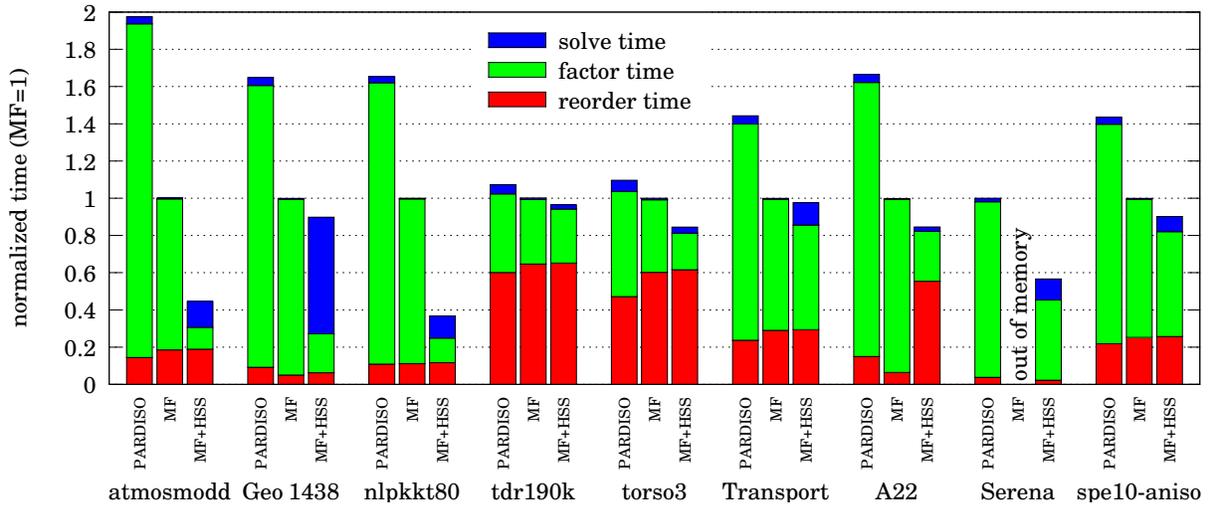}
  \end{center}    
  \caption{\footnotesize Comparison of timings for matrices from various
    applications on a $12$-core Intel\tm{} Ivy Bridge. PARDISO is the
    sparse direct multithreaded solver from Intel\tm{} MKL. MF refers
    to our implementation of the multifrontal method and MF+HSS is our
    new multifrontal solver with HSS compression. The matrices are
    taken from the Florida Sparse Matrix Collection and from SciDAC
    projects at the DOE. For these matrices, which are all quite large
    and from 2D/3D PDE problems, our MF solver is faster than PARDISO
    and HSS compression gives an additional speedup.}
  \label{fig:hist}
\end{figure}

Figure~\ref{fig:hist} shows a comparison of timings to solve linear
systems with a number of matrices from applications. The matrices
\textsc{atmosmodd}, \textsc{Geo\_1438}, \textsc{nlpkkt80},
\textsc{torso3}, \textsc{Transport} and \textsc{Serena} are from the
University of Florida Sparse Matrix
Collection\footnote{\url{http://www.cise.ufl.edu/research/sparse/matrices/}}. The
other matrices \textsc{tdr190k}, \textsc{A22} and
\textsc{spe10-anisotropic}, are from SciDAC projects at the DOE. The
matrices are also listed in Table~\ref{tab:matrix_list}, which
additionally contains matrix \textsc{Cube\_Coup\_dt0}. This last
matrix is not shown in Figure~\ref{fig:hist} because our multifrontal
code ran out of memory during factorization unless HSS compression was
used. All selected matrices are relatively large and originated from a
2D or 3D partial differential equation (on arbitrary domains). In
Figure~\ref{fig:hist}, our HSS enabled multifrontal solver (MF+HSS) is
compared to the pure multifrontal method (MF) and to the
state-of-the-art PARDISO solver~\cite{schenk2000efficient}. PARDISO, a
multithreaded supernodal solver, is part of Intel\tm{} MKL. For MF and
MF+HSS, reorder time includes nested dissection, MC64 and
symmetrization of the sparsity structure and for MF+HSS also separator
reordering. Factor time includes both symbolic and numerical
factorization. The times are normalized to a total time $1$ for
MF. For the matrices selected for Figure~\ref{fig:hist}, we see a
consistent speedup from MF+HSS compared to pure MF and our MF solver
always outperforms the PARDISO solver. PARDISO uses the same METIS
nested dissection reordering as our implementation, with comparable
reordering times for the different solvers. The supernodal pivoting
scheme used in PARDISO for numerical stability does not affect the
fill-in so the overall number of nonzeros in the factors with PARDISO
and with our multifrontal code are very similar. Only for the
\textsc{A22} problem reordering the separator in order to reduce the
HSS-ranks takes a lot of time. This is probably due to the addition of
link-two edges to the separator graph (see
Section~\ref{sec:sep_reorder}) since the original matrix already has
246 nonzeros per row on average. However, if those extra edges are not
taken into account, the HSS-ranks are much larger and there is no net
performance benefit from using HSS.

\begin{table}\footnotesize
  \renewcommand{\arraystretch}{1}
  \setlength{\tabcolsep}{0.20em}
  \newcommand{\ce}[1]{\multicolumn{1}{c|}{#1}}
  \newcommand{\cee}[1]{\multicolumn{1}{c|}{#1}}
  \begin{center}
    \begin{tabular}{|l||c|c||c|c||c|c|c|c|c|c|c|}
      \cline{4-11}
      \multicolumn{3}{c||}{} & \multicolumn{2}{|c||}{MF} & \multicolumn{6}{|c|}{HSS} \\
      \hline
      matrix     &  order & \#nnz & factor & solve  & $\ell_s / \ell_{\textrm{max}}$ & $\varepsilon$ & rank & its & factor & solve \\
      \hline  \hline
      atmosmodd  & 1.2M   & 8.8M    & 81s    & 0.4s   & 6/18  & $0.9$    & 17  & 88  & 25s   & 11s   \\
      \hline
      Geo\_1438  & 1.4M   & 63M   & 205s   & 1s     & 6/18  & $0.9$    & 8    & 318 & 56s   & 129s  \\
      \hline
      nlpkkt80   & 1.1M   & 29M   & 197s   & 0.7s   & 6/18  & $0.5$    & 59   & 90  & 49s   & 23s   \\
      \hline
      tdr190k    & 1.1M   & 43M   & 19s    & 0.2s   & 1/18  & $10^{-4}$ & 61   &  2 & 18s   & 0.4s  \\
      \hline
      torso3     & .25M   & 4.4M    & 6s     & 0.05s  & 6/15  & $0.5$   & 36  &  7 & 5s    & 0.2s  \\
      \hline
      Transport  & 1.6M   & 23M   & 80s    & 0.5s   & 3/18  & $10^{-2}$ & 182  & 24 & 69s   & 10s   \\
      \hline
      A22        & .59M   & 145M  & 127s   & 0.7s   & 10/17 & $0.1$    & 172  & 18 & 105s  & 3s    \\
      \hline
      spe10-aniso & 1.2M  & 31M   & 88s    & 0.4s   & 3/18  & $10^{-2}$ & 245  & 21 & 73s  & 7.3s    \\
      \hline
      Serena*     & 1.4M  & 65M   & 171s   & 0.5s   & 6/18  & $0.9$    & 11   & 111  & 40s  & 22s    \\
      \hline
      Cube\_Coup\_dt0* & 2.2M & 65M & -     & -      & 8/19  & $0.5$    & 100  & 200  & 60s  & 63s \\
      \hline
      \multicolumn{3}{l}{*single precision experiment}
    \end{tabular}
  \end{center}
  \caption{\footnotesize Same as in Figure~\ref{fig:hist}: comparison of timings for matrices from various
    applications on a $12$-core Intel\tm{} Ivy Bridge. This table also shows the optimal 
    number of HSS levels $\ell_s$, the optimal compression tolerance $\varepsilon$ and
    the corresponding HSS-rank and number of GMRES($30$) iterations.
    For the \textsc{Serena} and \textsc{Cube\_Coup\_dt0} matrices the pure multifrontal
    method ran out of memory in double precision. }
  \label{tab:matrix_list}
\end{table}

\subsection{Many-core parallel performance}
\begin{figure}
  \begin{center}
    \begin{subfigure}{.49\textwidth}
      \includegraphics[width=\textwidth]{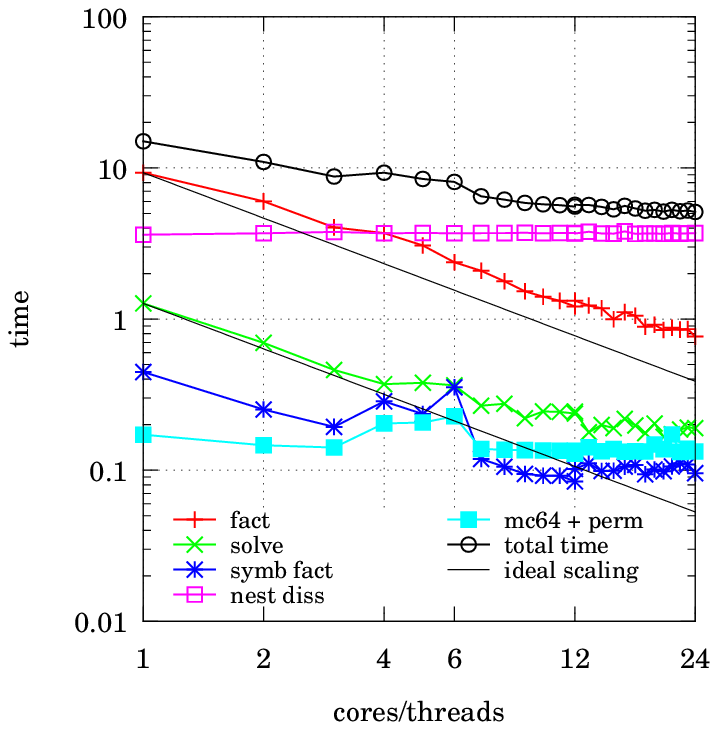}
      \caption{\footnotesize Intel\tm{} Ivy Bridge.}
      \label{fig:622}
    \end{subfigure}
    \begin{subfigure}{.49\textwidth}
      \includegraphics[width=\textwidth]{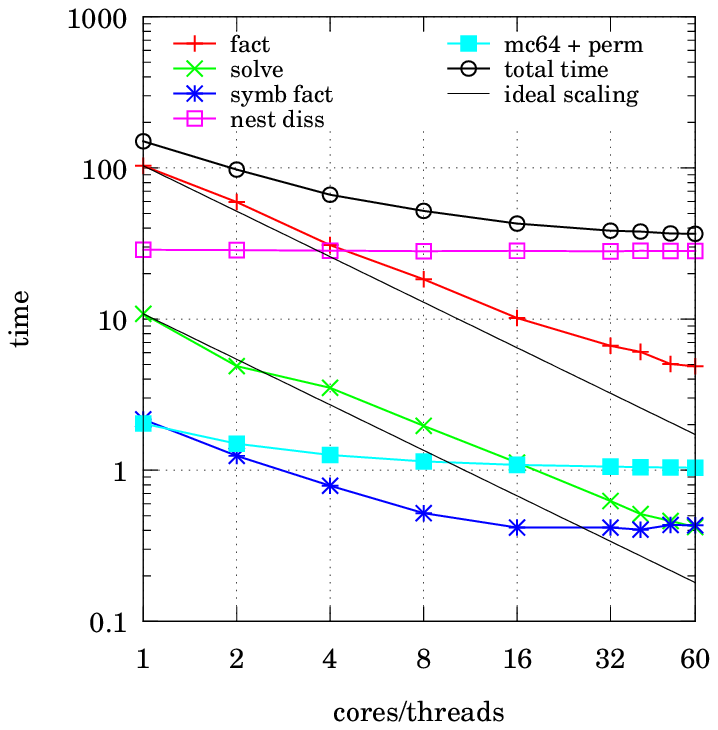}
      \caption{\footnotesize Intel\tm{} Xeon Phi.}
      \label{fig:623}
    \end{subfigure}
  \end{center}
  \caption{\footnotesize Multi-core scalability of the different steps in the MF+HSS
    solver on two leading architectures. The MF+HSS solver is applied
    to the relatively small \textsc{torso3.mtx} matrix. The code
    achieves good speedup for the numerical factorization phase and
    reasonable speedup for the solve (MF+HSS preconditioned
    GMRES). Note that the sequential reordering codes METIS and MC64
    become bottlenecks.}
  \label{fig:scaling_edison_babbage}
\end{figure}

Figure~\ref{fig:scaling_edison_babbage} shows performance and parallel
scalability of the MF+HSS solver applied to the \textsc{torso3.mtx}
matrix ($\ell_s=6$, $\varepsilon=0.5$) on two leading multi-core
architectures: a two sockets machine with a $12$-core Intel\tm{} Ivy
Bridge Xeon per socket and a $60$-core Intel\tm{} Xeon Phi Knight's
Corner. When running $12$ or less threads on the dual socket $24$-core
Xeon system, the threads are all bound to a single socket (NUMA
node). Note that since the Xeon Phi only has $8$GB of memory, the
larger problems from Table~\ref{tab:matrix_list} do not fit in it's
memory. Our code shows good parallel scalability on both architectures
for the numerical factorization phase and reasonable scalability for
the solve phase. However, with increasing number of threads the
reordering codes MC64 and METIS/SCOTCH quickly become scaling
bottlenecks. The MC64 phase in Figure~\ref{fig:scaling_edison_babbage}
shows some parallel speedup since this time also includes applying the
column permutation from MC64, which is done in parallel.

\section{Conclusions \& Outlook}\label{sec:conclusions}
We presented an initial attempt to create a high performance
implementation of a novel multifrontal solver with HSS low-rank
structures. We show speedups of up to $7\times$ over the pure
multifrontal algorithm for a range of applications. Moreover, our
implementation compares favorably to the commercial PARDISO solver. We
observed that the new solver has lower computational complexity than
the pure multifrontal method. However, the constants involved are much
larger. We will focus our attention on trying to reduce these
constants (for instance by trying to reduce the HSS-ranks) and on
solving larger problems with a distributed memory implementation. As
possible strategies to reduce the HSS-ranks we consider the
following. A power iteration on the random vectors, for instance $S^r
= \left( A A^* \right)^q A R$ with $q$ a small integer, will improve
the quality of the samples at the expense of additional computations;
see~\cite{halko2011finding} for further details. We believe the
separator reordering, see Section~\ref{sec:sep_reorder}, can be
improved, perhaps by taking into account the matrix entries and/or the
underlying geometry, also leading to lower ranks. Finally, a better
rank-revealing factorization, like a \emph{strong} rank-revealing
QR~\cite{gu1996efficient}, might lead to lower ranks and possibly more
stable ULV factorization. The solver with HSS compression achieves
lower floating point operation throughput than the pure multifrontal
code. Hence, we believe there is some room for improvement. We will
continue performance tuning of the code on various modern computer
architectures.

The presented code is part of a package called STRUMPACK. At the
moment STRUMPACK has a sparse shared memory solver and a dense
distributed memory solver. The longer term goal is to develop and
maintain a single scalable code for both sparse and dense problems
using hybrid parallelism. The current paper, together with the
distributed HSS code developed for~\cite{FHR} are a good step towards
reaching that goal.

The research on fast sparse and dense direct solvers is a very active
field at the moment. Some newer algorithmic ideas are for instance
nested HSS approximation and matrix-free direct solver-based
preconditioners. In nested HSS approximation, the HSS generators of
the frontal matrices are themselves HSS matrices. This could further
reduce the overall complexity of the solver. A matrix-free direct
solver based preconditioner could be constructed using randomization
techniques. It seems that knowledge of the sparsity pattern would be
required for this.

\section*{Acknowledgments}
We would like to thank Jianlin Xia for insightful discussions and for
his pioneering work on these exciting algorithms. Also thanks to Alex
Druinsky for testing an early version of the code. Yvan Notay provided
the code to generate the convection diffusion matrices and the code
for the Helmholtz problems was provided by Shen Wang.
% \todo{contact them and ask about code!}
% DOE funding
  Partial support for this work was provided through Scientific
  Discovery through Advanced Computing (SciDAC) program funded by
  U.S. Department of Energy, Office of Science, Advanced Scientific
  Computing Research (and Basic Energy Sciences/Biological and
  Environmental Research/High Energy Physics/Fusion Energy
  Sciences/Nuclear Physics).
% NERSC computing resources
  This research used resources of the National Energy Research
  Scientific Computing Center, which is supported by the Office of
  Science of the U.S. Department of Energy under Contract
  No. DE-AC02-05CH11231.

\bibliographystyle{plain}
\bibliography{refs}

\begin{thebibliography}{10}

\bibitem{agullo2013multifrontal}
Emmanuel Agullo, Alfredo Buttari, Abdou Guermouche, and Florent Lopez.
\newblock Multifrontal {QR} factorization for multicore architectures over
  runtime systems.
\newblock In {\em Euro-Par 2013 Parallel Processing}, pages 521--532. Springer,
  2013.

\bibitem{agullo2009numerical}
Emmanuel Agullo, Jim Demmel, Jack Dongarra, Bilel Hadri, Jakub Kurzak, Julien
  Langou, Hatem Ltaief, Piotr Luszczek, and Stanimire Tomov.
\newblock Numerical linear algebra on emerging architectures: The plasma and
  magma projects.
\newblock In {\em Journal of Physics: Conference Series}, volume 180, page
  012037. IOP Publishing, 2009.

\bibitem{SivaramPHD}
Sivaram Ambikasaran.
\newblock {\em {Fast Algorithms for Dense Numerical Linear Algebra and
  Applications}}.
\newblock PhD thesis, Stanford, 2013.

\bibitem{amestoy2014improving}
Patrick~R Amestoy, Cleve Ashcraft, Olivier Boiteau, Alfredo Buttari, Jean-Yves
  L'Excellent, and Cl\'{e}ment Weisbecker.
\newblock Improving multifrontal methods by means of block low-rank
  representations.
\newblock {\em To appear in SIAM Journal on Scientific Computing}, 2015.

\bibitem{amestoy1996amd}
Patrick~R Amestoy, Tim~A Davis, and Iain~S Duff.
\newblock {An approximate minimum degree ordering algorithm}.
\newblock {\em {SIAM Journal on Matrix Analysis and Applications}},
  17(4):886--905, 1996.

\bibitem{amestoy2001fully}
Patrick~R Amestoy, Iain~S Duff, Jean-Yves L'Excellent, and Jacko Koster.
\newblock A fully asynchronous multifrontal solver using distributed dynamic
  scheduling.
\newblock {\em SIAM Journal on Matrix Analysis and Applications}, 23(1):15--41,
  2001.

\bibitem{amestoy2008parallel}
Patrick~R Amestoy, Iain~S Duff, Daniel Ruiz, and Bora U{\c{c}}ar.
\newblock A parallel matrix scaling algorithm.
\newblock In {\em High Performance Computing for Computational Science-VECPAR
  2008}, pages 301--313. Springer, 2008.

\bibitem{AminfarAD14}
Amirhossein Aminfar, Sivaram Ambikasaran, and Eric Darve.
\newblock A fast block low-rank dense solver with applications to
  finite-element matrices.
\newblock {\em CoRR}, abs/1403.5337, 2014.

\bibitem{augonnet2011starpu}
C{\'e}dric Augonnet, Samuel Thibault, Raymond Namyst, and Pierre-Andr{\'e}
  Wacrenier.
\newblock {StarPU: a unified platform for task scheduling on heterogeneous
  multicore architectures}.
\newblock {\em Concurrency and Computation: Practice and Experience},
  23(2):187--198, 2011.

\bibitem{bebendorf2000approximation}
Mario Bebendorf.
\newblock Approximation of boundary element matrices.
\newblock {\em Numerische Mathematik}, 86(4):565--589, 2000.

\bibitem{blumofe1996cilk}
Robert~D Blumofe, Christopher~F Joerg, Bradley~C Kuszmaul, Charles~E Leiserson,
  Keith~H Randall, and Yuli Zhou.
\newblock Cilk: An efficient multithreaded runtime system.
\newblock {\em Journal of parallel and distributed computing}, 37(1):55--69,
  1996.

\bibitem{blumofe1999scheduling}
Robert~D Blumofe and Charles~E Leiserson.
\newblock Scheduling multithreaded computations by work stealing.
\newblock {\em Journal of the ACM (JACM)}, 46(5):720--748, 1999.

\bibitem{borm2003introduction}
Steffen B{\"o}rm, Lars Grasedyck, and Wolfgang Hackbusch.
\newblock Introduction to hierarchical matrices with applications.
\newblock {\em Engineering Analysis with Boundary Elements}, 27(5):405--422,
  2003.

\bibitem{bosilca2012dague}
George Bosilca, Aurelien Bouteiller, Anthony Danalis, Thomas Herault, Pierre
  Lemarinier, and Jack Dongarra.
\newblock Dague: A generic distributed dag engine for high performance
  computing.
\newblock {\em Parallel Computing}, 38(1):37--51, 2012.

\bibitem{buttari2009class}
Alfredo Buttari, Julien Langou, Jakub Kurzak, and Jack Dongarra.
\newblock A class of parallel tiled linear algebra algorithms for multicore
  architectures.
\newblock {\em Parallel Computing}, 35(1):38--53, 2009.

\bibitem{chan1987rank}
Tony~F Chan.
\newblock {Rank revealing {QR} factorizations}.
\newblock {\em Linear Algebra and Its Applications}, 88:67--82, 1987.

\bibitem{chandrasekaran2010numerical}
Shivkumar Chandrasekaran, Patrick Dewilde, Ming Gu, and N~Somasunderam.
\newblock {On the numerical rank of the off-diagonal blocks of Schur
  complements of discretized elliptic PDEs}.
\newblock {\em SIAM Journal on Matrix Analysis and Applications},
  31(5):2261--2290, 2010.

\bibitem{chandrasekaran2006fast}
Shivkumar Chandrasekaran, Ming Gu, and Timothy Pals.
\newblock {A fast ULV decomposition solver for hierarchically semiseparable
  representations}.
\newblock {\em SIAM Journal on Matrix Analysis and Applications},
  28(3):603--622, 2006.

\bibitem{cheng2005compression}
Hongwei Cheng, Zydrunas Gimbutas, Per-Gunnar Martinsson, and Vladimir Rokhlin.
\newblock On the compression of low rank matrices.
\newblock {\em SIAM Journal on Scientific Computing}, 26(4):1389--1404, 2005.

\bibitem{curtis1972automatic}
A~R Curtis and John~Ker Reid.
\newblock {On the automatic scaling of matrices for Gaussian elimination}.
\newblock {\em IMA Journal of Applied Mathematics}, 10(1):118--124, 1972.

\bibitem{recursiveTileLU}
Jack Dongarra, Mathieu Faverge, Hatem Ltaief, and Piotr Luszczek.
\newblock Achieving numerical accuracy and high performance using recursive
  tile lu factorization with partial pivoting.
\newblock {\em Concurrency and Computation: Practice and Experience},
  26(7):1408--1431, 2014.

\bibitem{duff1999design}
Iain~S Duff and Jacko Koster.
\newblock The design and use of algorithms for permuting large entries to the
  diagonal of sparse matrices.
\newblock {\em SIAM Journal on Matrix Analysis and Applications},
  20(4):889--901, 1999.

\bibitem{duff1983multifrontal}
Iain~S Duff and John~Ker Reid.
\newblock The multifrontal solution of indefinite sparse symmetric linear.
\newblock {\em ACM Transactions on Mathematical Software (TOMS)},
  9(3):302--325, 1983.

\bibitem{duran2011ompss}
Alejandro Duran, Eduard Ayguad{\'e}, Rosa~M Badia, Jes{\'u}s Labarta, Luis
  Martinell, Xavier Martorell, and Judit Planas.
\newblock Ompss: a proposal for programming heterogeneous multi-core
  architectures.
\newblock {\em Parallel Processing Letters}, 21(02):173--193, 2011.

\bibitem{eng11}
B.~Engquist and L.~Ying.
\newblock Sweeping preconditioner for the helmholtz equation: hierarchical
  matrix representation.
\newblock {\em Commun. Pure Appl. Math.}, 64:697--735, 2011.

\bibitem{frigo1999cache}
Matteo Frigo, Charles~E Leiserson, Harald Prokop, and Sridhar Ramachandran.
\newblock Cache-oblivious algorithms.
\newblock In {\em Foundations of Computer Science, 1999. 40th Annual Symposium
  on}, pages 285--297. IEEE, 1999.

\bibitem{ghemawat2009tcmalloc}
Sanjay Ghemawat and Paul Menage.
\newblock {Tcmalloc: Thread-caching malloc}.
\newblock {\em goog-perftools. sourceforge. net/doc/tcmalloc. html}, 2009.

\bibitem{gu1996efficient}
Ming Gu and Stanley~C Eisenstat.
\newblock {Efficient algorithms for computing a strong rank-revealing {QR}
  factorization}.
\newblock {\em SIAM Journal on Scientific Computing}, 17(4):848--869, 1996.

\bibitem{halko2011finding}
Nathan Halko, Per-Gunnar Martinsson, and Joel~A Tropp.
\newblock {Finding structure with randomness: Probabilistic algorithms for
  constructing approximate matrix decompositions}.
\newblock {\em SIAM review}, 53(2):217--288, 2011.

\bibitem{henon2002pastix}
Pascal H{\'e}non, Pierre Ramet, and Jean Roman.
\newblock {PaStiX: a high-performance parallel direct solver for sparse
  symmetric positive definite systems}.
\newblock {\em Parallel Computing}, 28(2):301--321, 2002.

\bibitem{hogg2010design}
Jonathan~D Hogg, John~K Reid, and Jennifer~A Scott.
\newblock {Design of a multicore sparse Cholesky factorization using DAGs}.
\newblock {\em SIAM Journal on Scientific Computing}, 32(6):3627--3649, 2010.

\bibitem{karypis1998fast}
George Karypis and Vipin Kumar.
\newblock A fast and high quality multilevel scheme for partitioning irregular
  graphs.
\newblock {\em SIAM Journal on scientific Computing}, 20(1):359--392, 1998.

\bibitem{kim2014parallel}
Kyungjoo Kim and Victor Eijkhout.
\newblock {A parallel sparse direct solver via hierarchical DAG scheduling}.
\newblock {\em ACM Transactions on Mathematical Software (TOMS)}, 41(1):3,
  2014.

\bibitem{kriemann2013}
Ronald Kriemann.
\newblock {$\mathcal{H}$-{LU} factorization on many-core systems}.
\newblock {\em Computing and Visualization in Science}, 16(3):105--117, 2013.

\bibitem{lacoste2014taking}
Xavier Lacoste, Mathieu Faverge, George Bosilca, Pierre Ramet, and Samuel
  Thibault.
\newblock Taking advantage of hybrid systems for sparse direct solvers via
  task-based runtimes.
\newblock In {\em Parallel \& Distributed Processing Symposium Workshops
  (IPDPSW), 2014 IEEE International}, pages 29--38. IEEE, 2014.

\bibitem{lexcellent2012multifrontal}
Jean-Yves L'Excellent.
\newblock {\em Multifrontal Methods: Parallelism, Memory Usage and Numerical
  Aspects}.
\newblock Habilitation \`a diriger des recherches, \'Ecole normale sup\'erieure
  de Lyon, September 2012.

\bibitem{lin2011fast}
Lin Lin, Jianfeng Lu, and Lexing Ying.
\newblock Fast construction of hierarchical matrix representation from
  matrix--vector multiplication.
\newblock {\em Journal of Computational Physics}, 230(10):4071--4087, 2011.

\bibitem{liu1992multifrontal}
Joseph W.~H. Liu.
\newblock The multifrontal method for sparse matrix solution: Theory and
  practice.
\newblock {\em SIAM review}, 34(1):82--109, 1992.

\bibitem{martinsson2011fast}
Per-Gunnar Martinsson.
\newblock A fast randomized algorithm for computing a hierarchically
  semiseparable representation of a matrix.
\newblock {\em SIAM Journal on Matrix Analysis and Applications},
  32(4):1251--1274, 2011.

\bibitem{streambench}
John~D McCalpin.
\newblock {STREAM: Sustainable Memory Bandwidth in High Performance Computers}.
\newblock http://www.cs.virginia.edu/stream/.

\bibitem{mccool2012structured}
Michael McCool, James Reinders, and Arch Robison.
\newblock {\em Structured parallel programming: patterns for efficient
  computation}.
\newblock Elsevier, 2012.

\bibitem{napov2013algebraic}
Artem Napov and Xiaoye~S Li.
\newblock An algebraic multifrontal preconditioner that exploits the low-rank
  property.
\newblock {\em Submitted to Numerical Linear Algebra with Applications}, 2013.

\bibitem{agmg}
Y.~Notay.
\newblock An aggregation-based algebraic multigrid method.
\newblock {\em Electronic Trans. Numer. Anal.}, 37:123--146, 2010.

\bibitem{pellegrini1996scotch}
Fran{\c{c}}ois Pellegrini and Jean Roman.
\newblock Scotch: A software package for static mapping by dual recursive
  bipartitioning of process and architecture graphs.
\newblock In {\em High-Performance Computing and Networking}, pages 493--498.
  Springer, 1996.

\bibitem{quintana1998blas}
Gregorio Quintana-Ort{\'\i}, Xiaobai Sun, and Christian~H Bischof.
\newblock {A BLAS-3 version of the {QR} factorization with column pivoting}.
\newblock {\em SIAM Journal on Scientific Computing}, 19(5):1486--1494, 1998.

\bibitem{reinders2007intel}
James Reinders.
\newblock {\em {Intel threading building blocks: outfitting C++ for multi-core
  processor parallelism}}.
\newblock O'Reilly Media, Inc., 2007.

\bibitem{FHR}
Fran\c{c}ois-Henry Rouet, Xiaoye~S. Li, Pieter Ghysels, and Artem Napov.
\newblock {A distributed-memory package for dense Hierarchically Semi-Separable
  matrix computations using randomization}.
\newblock {\em submitted to ACM Transactions on Mathematical Software}, 2014.

\bibitem{ruiz2001scaling}
Daniel Ruiz.
\newblock A scaling algorithm to equilibrate both rows and columns norms in
  matrices.
\newblock Technical report, Technical Report RT/APO/01/4, ENSEEIHT-IRIT, 2001.
  Also appeared as RAL report RAL-TR-2001-034, 2001.

\bibitem{schenk2000efficient}
Olaf Schenk, Klaus G{\"a}rtner, and Wolfgang Fichtner.
\newblock Efficient sparse lu factorization with left-right looking strategy on
  shared memory multiprocessors.
\newblock {\em BIT Numerical Mathematics}, 40(1):158--176, 2000.

\bibitem{vandebril2005bibliography}
Raf Vandebril, Marc Van~Barel, Gene Golub, and Nicola Mastronardi.
\newblock A bibliography on semiseparable matrices.
\newblock {\em Calcolo}, 42(3-4):249--270, 2005.

\bibitem{wang20113d}
Shen Wang, Maarten~V de~Hoop, and Jianlin Xia.
\newblock {On 3D modeling of seismic wave propagation via a structured parallel
  multifrontal direct Helmholtz solver}.
\newblock {\em Geophysical Prospecting}, 59(5):857--873, 2011.

\bibitem{wang2014parallel}
Shen Wang, Xiaoye~S Li, Fran\c{c}ois-Henry Rouet, Jianlin Xia, and Maarten~V
  De~Hoop.
\newblock A parallel geometric multifrontal solver using hierarchically
  semiseparable structure.
\newblock {\em submitted to ACM Transactions on Mathematical Software}, 2014.

\bibitem{weisbecker2013improving}
Cl{\'e}ment Weisbecker.
\newblock {\em {Improving multifrontal solvers by means of algebraic Block
  Low-Rank representations}}.
\newblock PhD thesis, Institut National Polytechnique de Toulouse, 2013.

\bibitem{wilkinson1994rounding}
James~Hardy Wilkinson.
\newblock {\em Rounding errors in algebraic processes}.
\newblock Courier Dover Publications, 1994.

\bibitem{xia2013randomized}
Jianlin Xia.
\newblock Randomized sparse direct solvers.
\newblock {\em SIAM Journal on Matrix Analysis and Applications},
  34(1):197--227, 2013.

\bibitem{xia2010fast}
Jianlin Xia, Shivkumar Chandrasekaran, Ming Gu, and Xiaoye~S Li.
\newblock Fast algorithms for hierarchically semiseparable matrices.
\newblock {\em Numerical Linear Algebra with Applications}, 17(6):953--976,
  2010.

\bibitem{xia2012superfast}
Jianlin Xia, Yuanzhe Xi, and Ming Gu.
\newblock {A superfast structured solver for Toeplitz linear systems via
  randomized sampling}.
\newblock {\em SIAM Journal on Matrix Analysis and Applications},
  33(3):837--858, 2012.

\bibitem{yarkhan2011quark}
Asim YarKhan, Jakub Kurzak, and Jack Dongarra.
\newblock Quark users’ guide: Queueing and runtime for kernels.
\newblock {\em University of Tennessee Innovative Computing Laboratory
  Technical Report ICL-UT-11-02}, 2011.

\end{thebibliography}

\end{document}